\documentclass[12pt]{article}
\usepackage{array,amsmath,amssymb,epsfig,verbatim}
\begin{document}


\newcommand{\kIR}{K_{\Lambda\infty}}
\newcommand{\kUV}{K_{0\Lambda}}

\def\slash#1{\ooalign{$\hfil/\hfil$\crcr$#1$}}
\def\KM#1{K_{\Lambda\Lambda_0}(-(#1)^2)} 

\newcommand{\bracket}[2]
{
\left[ #1 \right]_{#2}
}
\newcommand{\sezione}[1]
{
{$\ $ \\ \bf\large #1 \\ } \addcontentsline{toc}{section}{#1}
}


\def\ap#1#2#3{Ann.\ Phys.\ (NY) #1 (19#3) #2}
\def\cmp#1#2#3{Commun.\ Math.\ Phys.\ #1 (19#3) #2}
\def\ib#1#2#3{ibid.\ #1 (19#3) #2}
\def\zp#1#2#3{Z.\ Phys.\ #1 (19#3) #2}
\def\np#1#2#3{Nucl.\ Phys.\ B#1 (19#3) #2}
\def\pl#1#2#3{Phys.\ Lett.\ #1B (19#3) #2}
\def\pr#1#2#3{Phys.\ Rev.\ D #1 (19#3) #2}
\def\prb#1#2#3{Phys.\ Rev.\ B #1 (19#3) #2}
\def\prep#1#2#3{Phys.\ Rep.\ #1 (19#3) #2}
\def\prl#1#2#3{Phys.\ Rev.\ Lett.\ #1 (19#3) #2}
\def\rmp#1#2#3{Rev.\ Mod.\ Phys.\ #1 (19#3) #2}

\newcommand{\figura}[4]
{
\begin{figure}[t]
  \begin{center}
  \mbox{\epsfig{file=#1,height=#2}}
  \end{center}
  \label{#3}
  \caption{{{\small #4}}}
\end{figure}
}
\newcommand{\figuraX}[2]
{
\begin{figure}[h]
  \begin{center}
  \mbox{\epsfig{file=#1,height=#2}}
  \end{center}
\end{figure}
}
\renewcommand{\div}
{
\mbox{div}
}
\newcommand{\binomio}[2]
{
\pmatrix{#1\cr #2}
}
\newcommand{\disegno}[2]
{
\begin{center}                             
  \begin{picture}(#1) #2
  \end{picture}
\end{center}
}
\newcommand{\vettore}[3]
{
\put(#1){\vector(#2){#3}}
}
\newcommand{\entra}[4]
{
\put(#1){\makebox(0,0){$\boldmath\times$}}
\put(#1){\line(#2){#3}}
\put(#1){\makebox(0,-4){$#4$}}
}
\newcommand{\entraX}[4]
{
\put(#1){\makebox(0,0){$\boldmath\times$}}
\put(#1){\line(#2){#3}}
\put(#1){\makebox(-3,-2){$#4$}}
}
\newcommand{\scrivi}[2]
{
\put(#1){\makebox(0,0){$#2$}}
}
\newcommand{\blob}[2]
{
\put(#1){\circle{2}}
\put(#1){\makebox(-.5,-.5){$#2$}} 
}                    
\newcommand{\croce}[2]
{
\put(#1){\makebox(0,0){$\boldmath\times$}}
\put(#1){\makebox(-3,-2){$#2$}} 
}                       
\newcommand{\puntino}[2]
{
\put(#1){\circle*{.5}}
\put(#1){\makebox(0,-4){$#2$}} 
}                 
\newcommand{\puntinoX}[2]
{
\put(#1){\circle*{.5}}
\put(#1){\makebox(-3,-2){$#2$}} 
}                       
\newcommand{\punto}[2]
{
\put(#1){\circle*{.9}}
\put(#1){\makebox(0,-4){$#2$}} 
}
\newcommand{\puntoX}[2]
{
\put(#1){\circle*{.9}}
\put(#1){\makebox(-3,-2){$#2$}} 
}
\newcommand{\puntone}[2]
{
\put(#1){\circle*{1.5}}
\put(#1){\makebox(0,-4){$#2$}} 
}
\newcommand{\puntoneX}[2]
{
\put(#1){\circle*{1.5}}
\put(#1){\makebox(-3,-2){$#2$}} 
}
\newcommand{\acapo}
{
\\ \ \\ 
}
\newcommand{\HUGE}
{
 \Huge\bf
}
\newcommand{\CENTRA}[1]
{
\begin{center} {\Huge\bf\underbar{#1}} \end{center}
}
\newcommand{\CENtra}[1]
{
\begin{center} {\huge\bf #1} \end{center}
}
\newcommand{\Centra}[1]
{
\begin{center} {\Large\bf #1} \end{center}
}
\newcommand{\centra}[1]
{
\begin{center} {\bf #1} \end{center}
}
\newcommand{\gr}[1]
{
^{\underline{#1}} 
}
\newcommand{\grr}[2]
{
^{\underline{#1\times #2}} 
}
\newcommand{\des}[1]
{
\overleftarrow{#1}
}
\newcommand{\ddes}[2]
{
{\overleftarrow\delta #1\over\delta #2}
}
\newcommand{\dsin}[2]
{
{\overrightarrow\delta #1\over\delta #2}
}
\newcommand{\valuta}[1]
{
\left.#1\right|
}

\newcommand{\iacc}{\`\i{\ }}
\renewcommand{\a}{{\tilde a}} \renewcommand{\b}{{\tilde b}}
\renewcommand{\c}{{\tilde c}} \renewcommand{\d}{{\tilde d}}
\newcommand{\e}{{\tilde e}}\newcommand{\f}{{\tilde f}}
\newcommand{\g}{{\tilde g}}\newcommand{\h}{{\tilde h}}
\newcommand{\ii}{{\tilde \imath}} \newcommand{\jj}{{\tilde \jmath}}
\renewcommand{\k}{{\tilde k}} \renewcommand{\l}{{\tilde l}}
\newcommand{\m}{{\tilde m}} \newcommand{\n}{{\tilde n}}
\renewcommand{\o}{{\tilde o}} \newcommand{\p}{{\tilde p}}
\newcommand{\q}{{\tilde q}} \renewcommand{\r}{{\tilde r}}
\newcommand{\s}{{\tilde s}} \renewcommand{\t}{{\tilde t}}
\renewcommand{\u}{{\tilde u}} \renewcommand{\v}{{\tilde v}}
\newcommand{\w}{{\tilde w}} \newcommand{\x}{{\tilde x}}
\newcommand{\y}{{\tilde y}} \newcommand{\z}{{\tilde z}}
\newcommand{\ux}{{\underline x}} \newcommand{\uy}{{\underline y}}
\newcommand{\uv}{{\underline v}} \newcommand{\uw}{{\underline w}}
\newcommand{\uz}{{\underline z}}
\newcommand{\bi}{\bar\imath} \newcommand{\bj}{\bar\jmath}


\newcommand{\tadpole}{\mbox{ 
\begin{picture}(12,4)
  \put(0,1){\line(1,0){10}}
  \put(5,2.5){\circle{3}}
\end{picture}}}

\newcommand{\zetadue}{\mbox{
\begin{picture}(6,3)
  \put(0,1){\line(1,0){6}}
  \put(3,1){\circle{2}}
\end{picture}}}

\newcommand{\vertice}{\mbox{ 
\begin{picture}(14,6)
 \put(6,2){\circle{8}}
 \put(2,2){\line(-1,1){4}}
 \put(2,2){\line(-1,-1){4}}
 \put(10,2){\line(1,1){4}}
 \put(10,2){\line(1,-1){4}}
\end{picture}}}

\newcommand{\verticeD}{\mbox{ 
\begin{picture}(8,3)
  \put(1,2){\line(1,0){6}}
  \put(4,2){\circle{2}}
  \put(4,1){\line(-1,-1){1.5}}
  \put(4,1){\line(1,-1){1.5}}
\end{picture}}}

\newcommand{\doppiovert}{\mbox{ 
\begin{picture}(6,3)
  \put(1,1){\line(-1,1){1.5}}
  \put(1,1){\line(-1,-1){1.5}}
  \put(2,1){\circle{2}}
  \put(4,1){\circle{2}}
  \put(5,1){\line(1,1){1.5}}
  \put(5,1){\line(1,-1){1.5}}
\end{picture}}}

\newcommand{\doppioquatt}{\mbox{ 
\begin{picture}(6,3)
  \put(2,1){\line(-1,1){1.5}}
  \put(2,1){\line(-1,-1){1.5}}
  \puntino{5,1}{}
  \put(3.5,1){\circle{3}}
  \put(5,1){\line(1,1){1.5}}
  \put(5,1){\line(1,-1){1.5}}
\end{picture}}}

\newcommand{\verticesei}{\mbox{
\begin{picture}(14,6)
 \put(6,2){\circle{8}}
 \put(6,6){\line(1,1){4}}
 \put(6,6){\line(-1,1){4}}
 \put(2,2){\line(-1,1){4}}
 \put(2,2){\line(-1,-1){4}}
 \put(10,2){\line(1,1){4}}
 \put(10,2){\line(1,-1){4}}
\end{picture}}}

\newcommand{\ctdue}{\mbox{
\begin{picture}(12,2)
  \put(0,1){\line(1,0){10}} \puntone{5,1}{}
\end{picture}}}

\newcommand{\ctquatt}{\mbox{
\begin{picture}(12,2)
  \put(0,1){\line(1,0){10}}
  \puntone{5,1}{}
  \put(5,1){\line(1,-1){4}}
  \put(5,1){\line(-1,-1){4}}
\end{picture}}}
\def\bom#1{\mbox{\boldmath$#1$}}

\def\ldl{\Lambda_0\frac{\partial}{\partial\Lambda_0}}
\def\mdm{\mu\frac\partial{\partial\mu}}
\def\lDl{\Lambda_0\frac{d}{d\Lambda_0}} 

\def\LdL{\Lambda\partial_\Lambda}
\def\LDL{\Lambda \delta_\Lambda}

\def\UV{ {\mbox{\scriptsize UV}} }
\def\limUV{\lim_{\Lambda_0\to\infty}}
\def\MSbar{ \overline{MS} }
\def\arctanh{\mbox{arctanh}}


\newcommand{\hor}
  {\mbox{hor}}
\newcommand{\ver}
  {\mbox{ver}}
\newcommand{\Tr}
  {\mbox{Tr}}
\newcommand{\STr}
  {\mbox{STr}}
\newcommand{\A}
  {{\cal A}}
\newcommand{\B}
  {{\cal B}}
\newcommand{\C}
  {{\cal C}}                                  
\newcommand{\D}
  {{\cal D}}
\newcommand{\De}
  {{\cal D}}
\newcommand{\E}
  {{\cal E}}
\newcommand{\F}
  {{\cal F}}
\newcommand{\G}
  {{\cal G}}
\renewcommand{\H}
  {{\cal H}}
\newcommand{\I}
  {{\cal I}}
\newcommand{\J}
  {{\cal J}}
\newcommand{\K}
  {{\cal K}}
\renewcommand{\L}
  {{\cal L}}
\newcommand{\M}
  {{\cal M}}
\newcommand{\N}
  {{\cal N}}
\renewcommand{\O}
  {{\cal O}}
\renewcommand{\P}
  {{\cal P}}
\newcommand{\Q}
  {{\cal Q}}
\newcommand{\R}
  {{\cal R}}
\renewcommand{\S}
  {{\cal S}}
\newcommand{\T}
  {{\cal T}}
\newcommand{\U}
  {{\cal U}}
\newcommand{\V}
  {{\cal V}}
\newcommand{\W}
  {{\cal W}}
\newcommand{\X}
  {{\cal X}}
\newcommand{\Y}
  {{\cal Y}}
\newcommand{\Z}
  {{\cal Z}}
\newcommand{\uno}
  {\unity}
\newcommand{\lie}
  {\ell_\varepsilon}
\newcommand{\lt}[1]
  { \ell_{t_{#1}} }
\newcommand{\leps}[1]
  { \ell_{\varepsilon_{#1}} }
\newcommand{\KG}
{\mbox{ \tiny \begin{tabular}{|c|} \hline \\ \hline \end{tabular} } }
\newcommand{\sedici}
{\overline{16}}

\newcommand{\alp}{{\dot\alpha}}
\newcommand{\bet}{{\dot\beta}}
\newcommand{\gamm}{{\dot\gamma}}
\newcommand{\delt}{{\dot\delta}}
\newcommand{\epsilo}{{\dot\epsilon}}
\newcommand{\et}{{\dot\eta}}
\newcommand{\iot}{{\dot\iota}}
\newcommand{\kapp}{{\dot\kappa}}
\newcommand{\lambd}{{\dot\lambda}}
\newcommand{\muu}{{\dot\mu}}
\newcommand{\nuu}{{\dot\nu}}
\newcommand{\sigm}{{\dot\sigma}}
\newcommand{\ta}{{\dot\tau}}
\newcommand{\omeg}{{\dot\omega}}
\newcommand{\ep}{\varepsilon}
\newcommand{\lamb}{\tilde\lambda}

\newcommand{\ansatz}{{\it Ans\"atze }}


\newcommand{\PropI}
{
\Delta^{-1}_{\Lambda\Lambda_0}
}
\newcommand{\Prop}
{
\Delta_{\Lambda\Lambda_0}
}
\newcommand{\rif}[1]
  {(\ref{#1})}
\newcommand{\ebarra}
  {$\bar e\ $}
\newcommand{\limPoincare}
  {$\bar e\longrightarrow 0\ $}
\newcommand{\inbasso}[1]
  { _{{\cal #1}} }
\newcommand{\inalto}[1]
  { ^{{\cal #1}} }
\newcommand{\call}[1]
  {{\cal #1}}
\newcommand{\eps}[1]
  {\varepsilon_{a_1 \cdots a_D}}
\newcommand{\xx}[2]
  {\frac {1}{2} #1_{aa'}#2^{aa'}}
\newcommand{\xxx}[2]
  {\frac 12 #1_{abc}#2^{abc}}
\newcommand{\xxxx}[3]
  { \frac12#2_{a_1\cdots a_{#1}}#3^{a_1\cdots a_{#1}} }
\newcommand{\formula}[2]
  { \begin{equation} \label{#1} #2 \end{equation} }
\newcommand{\formule}[2]
{  
  \begin{subequations} \label{#1}
  \begin{eqnarray} 
	#2
  \end{eqnarray} 
  \end{subequations}
}

\newcommand{\formulona}[2]
{
  \begin{equation} \label{#1}
  \begin{split}
    #2 
  \end{split}
  \end{equation}
}
\newcommand{\formulonaX}[1]
{
  \begin{equation} 
    \begin{split}
      #1
    \end{split}
  \end{equation}
}
\newcommand{\graffa}[2]
{
  \begin{equation} \label{#1}
    \left\{ \begin{array}{lll} #2 \end{array} \right.
  \end{equation}
}
\newcommand{\graffaetichettata}[3]
{
  \begin{equation} \label{#2}
    #1\ \left\{ \begin{array}{lll} #3 \end{array} \right.
  \end{equation}
}
\newcommand{\formulaX}[1]
{  \begin{equation} #1 \end{equation} 
}
\newcommand{\formuleX}[1]
{ 
  \begin{subequations} 
  \begin{eqnarray} 
	#1 
  \end{eqnarray} 
  \end{subequations}
}
\newcommand{\graff}[1]
{
  $$  \left\{ \begin{array}{lll} #1 \end{array} \right. $$
}
\newcommand{\graffet}[2]
{
 $$  #1\ \left\{ \begin{array}{lll} #2 \end{array} \right. $$
}
\newcommand{\graffaX}[1]
{
  \begin{equation} 
    \left\{ \begin{array}{lll} #1 \end{array} \right.
  \end{equation}
}
\newcommand{\graffaetichettataX}[2]
{
  \begin{equation}
    #1\ \left\{ \begin{array}{lll} #2 \end{array} \right.
  \end{equation}
}
\newcommand{\detizero}[1]
{
\left.{d\over dt}#1 \right|_{t=0}
}
\newcommand{\prendi}[2]
{
\left. #1 \right|_{#2}
}
\newcommand{\dezero}[2]
{
\left.{\mbox{d} #1\over \mbox{d} #2} \right|_{#2=0}
}
\newcommand{\desude}[1]
{
{\partial \over \partial #1}
}
\newcommand{\dxx}[2]
{
{ \partial x^{#1}\over\partial x^{#2} }
}
\newcommand{\dd}[2]
{
\frac{\delta{#1}}{\delta{#2}}
}
\newcommand{\dede}[2]
{
{\partial{#1}\over\partial{#2}}
}
\newcommand{\dedi}[2]
{
{\mbox{d}{#1}\over \mbox{d}{#2}}
}
\newcommand{\dexx}[2]
{
{ \partial x^{#1}\over\partial x^{#2} }
}
\newcommand{\dexy}[1]
{
{ \partial x^{#1}\over\partial x^{#1'} }
}
\newcommand{\deyx}[1]
{
{ \partial x^{#1'}\over\partial x^{#1} }
}


\begin{titlepage}
\renewcommand{\thefootnote}{\fnsymbol{footnote}}
\begin{flushright}
     LPTHE 98-08\\
     August 1998 \\
\end{flushright}
\par \vskip 10mm
\begin{center}
{\Large \bf
Gauge Consistent Wilson Renormalization Group I: Abelian Case}
\end{center}
\par \vskip 2mm
\begin{center}
{\large M.\ Simionato\footnote{E-mail: micheles@lpthe.jussieu.fr}}\\
\vskip 5 mm
{\small \it
LPTHE, Universit\'e Pierre et Marie Curie (Paris VI) et Denis Diderot 
(Paris VII), Tour 16, 1er. \'etage, 4, Place Jussieu, 75252 Paris, Cedex 05, 
France  and Istituto Nazionale di Fisica Nucleare, Frascati, Italy }
\end{center}
\par \vskip 2mm
\begin{center} {\large \bf Abstract} \end{center}
{\small \begin{quote}
A version of the Exact Renormalization Group Equation consistent with 
gauge symmetry is presented. A discussion of its regularization and
renormalization is given. The relation with 
the Callan-Symanzik equation is clarified.\\

  Pacs: 11.10.Hi, 11.15.-q, 11.15.Bt.\\
  Keywords: Renormalization Group, Ward identities, QED
\end{quote}}
\end{titlepage}

\section{Introduction}

The problem of the perturbative renormalization of a quantum
field theory (possibly with gauge symmetry) is highly non-trivial. 
Therefore even after its complete rigorous solution in the BPHZL formalism
in the seventies \cite{BPH,Zimmermann,Lowenstein} 
there has been a big effort in the literature
in order to find alternative and simpler approaches. In
particular in recent years the Wilsonian point of view 
\cite{Wilson,Wegner-H,Polchinski,Hasenfratz} has gained more and more
popularity and nowadays is commonly regarded by many theorists not only
as the better alternative to the traditional formulation of quantum field
theory, but maybe also as the {\it correct way of thinking} about quantum field
theory \cite{ZJ.book,Peskin,Weinberg.book}. 
The reasons of this
success are clear: first, the approach is very physically appealing;
second, it is well founded at the mathematical level.
In particular, in recent years, our understanding of the technical aspects
of the formalism is much improved; for instance the perturbative
implementation of symmetries has been clarified
\cite{Becchi} and the relation with the BPHZ approach has been
understood \cite{KK.OpComp}. Nevertheless, the practical implementation of 
the Wilsonian formalism suffers for a technical very annoying problem, i.e. the
explicit breaking of gauge-invariance. This is due to an inconsistency
between the Wilson's Exact Renormalization Group Equation (ERGE) and the 
Ward-Takahashi identities.
Therefore a lot of non-trivial
work is needed to recover gauge-invariance on physical quantities.
In particular the all perturbative
machinery of  Quantum Action Principle \cite{QAP,Piguet}
and fine-tuning conditions seems needed. This fact is very unpleasant, 
because
it is unclear why in theories like QED or QCD, where there are traditional 
renormalization methods explicitly consistent with the gauge symmetry, the
Wilsonian formalism should be so bad. In particular one could think that
in a theory as simple as QED should be possible to implement a consistent
Wilsonian formulation, at least at the perturbative
level. However, to the best of 
our knowledge, no such a formulation appeared in the literature. 
Here we fill this gap.

Our basic idea is simply of introducing the Wilsonian
infrared cutoff as a mass-like term for both photons and electrons:
in this way the basic structure of Ward identities
is preserved. This idea is quite natural, nevertheless its implementation
is not straightforward. The point is that the mass cutoff
does not sufficiently regularize the theory and
to be properly managed requires an intermediate ultraviolet
regularization of the evolution equation. In this paper 
we discuss in detail how to
perform this task which is non-usual in a Wilsonian context.
Our result open the door to the application of non-perturbative numerical 
approximation schemes consistent with Ward identities, thus giving a 
strong improvement with respect to the
methods currently used in the Wilsonian literature 
\cite{Freire,Ellw.pot,Jungnickel,giap.QED}.

The plan of the paper is the following: in section 2 we give a short
introduction about the Wilsonian point of view and the previous work 
on the gauge-invariance problem; in section 3 we fix the
notation for the quantum electrodynamics and we write down the
Ward identities; in section 4 we implement the Exact
Renormalization Group Equation in a form suitable for
the following analysis; in section 5 we briefly review how it can be 
perturbatively solved in the loop-wise expansion; in sections 6 and 7 some
explicit one-loop computation are presented. 
A general analysis of Ward identities is given in section 8;
in section 9 we explain how
to extract the Callan-Symanzik equation directly from the ERGE.
Section 10 contains our conclusions and the outlook: various possible
extensions and physical application of the formalism are suggested. 
In particular we stress
that the formulation is perfectly calculative: the framework should 
be considered not only useful to study formal questions, but also 
for practical purposes. 
Three appendices on technical questions close the paper.

\section{The Wilsonian point of view and the problem of gauge invariance}

We begin by summarizing the basics of the Wilsonian point of view, as needed
for the applications to quantum field theory.
\begin{enumerate}
\item The fundamental object of the formalism
is the effective action at the scale $\Lambda$,
obtained by integrating out the ultraviolet degrees of freedom.
\item The procedure of integrating degrees of freedom is converted
into the problem of solving a differential equation in $\Lambda$, the Wilson's
Exact Renormalization Group Equation. 
In this way, by the knowledge of the ultraviolet physics
(i.e. of the effective action at some ultraviolet scale 
$\Lambda=\Lambda_{UV}$) one can deduce the infrared physics (i.e. the 
effective action at some infrared scale $\Lambda=\Lambda_{IR}<<\Lambda_{UV}$) 
by solving the ERGE.
\item The infrared effective action is independent on the details of the
ultraviolet physics, i.e. it depends only on a little number of relevant
parameters. This is the physical meaning of the renormalizability 
(universality in statistical language) property.
\end{enumerate}

Points 1,2,3 are common to all the approaches based on the Wilson's 
point of view; however the various technical implementation of the formalism
are strongly author dependent and very different in practice. For instance the
degrees of freedom integration can be done {\it \`a la} Wegner-Houghton,
by integrating the momenta on a shell of thickness
$\delta\Lambda$, or {\it \`a la} Polchinski,
by introducing a smooth cutoff function which multiplies the free propagators
of the theory. Moreover, one can take as fundamental 
effective action the
Wilsonian action $S_{eff}(\Phi;\Lambda)$ or, alternatively, its Legendre
transform $\Gamma(\Phi;\Lambda)$ (sometimes called effective
average action \cite{Wetterich1} or simply effective cutoff action
\cite{BDM}). This latter formalism
is better suited for a comparison with the traditional renormalization
theory and will be adopted 
in this work. \\

As we said, all the usual formulation
of the evolution equation are inconsistent with gauge-invariance, thus
the flow does not preserve the symmetry: 
even if the ultraviolet action is gauge-invariant, the infrared is not.
Conversely, in order to have a gauge-invariant infrared action,
one is forced to start with a non-gauge-invariant ultraviolet action.
There was a big effort in the literature to face this problem. 
Here we give a short review of various solutions 
proposed in the past, with no pretense of completeness.
In particular we restrict ourself to the Polchinski's formulation 
of the evolution equation or its Legendre transformed version, neglecting
some work in other formalisms, as for instance \cite{Liao}.
\begin{enumerate}
\item Maybe the first attempt, following the original Polchinski
formalism, was the work of Warr \cite{Warr}. In this paper the idea is
of using an explicit gauge-invariant 
Pauli-Villars regularization supplemented by higher deri\-vative
terms. This idea is quite simple in principle, but in practice the rigorous
formulation is very technical and it needs as an intermediate step a 
pre-regulator, i.e. a momentum
cutoff, which explicitly breaks gauge-invariance; moreover 
concrete computations are difficult
to perform and, up to our knowledge, this approach never was pursued
in the successive literature.
\item A second very important point of view 
was advocated by  Becchi \cite{Becchi}.
In this approach the attention is on rigorous proofs concerning the
perturbative recovering of the symmetry for the physical objects. 
Put in other way, the Wilson 
Renormalization Group in the Polchinski implementation is used to
prove the Quantum Action Principle \cite{QAP} 
of perturbative quantum field theory. In this way it is possible to
show that the gauge symmetry can be recovered via a perturbative
fine tuning of a finite number of relevant couplings, provided that 
the theory is anomaly free.
Unfortunately, the explicit solution of the fine-tuning conditions is
extremely cumbersome beyond one-loop, even in simple models 
\cite{FerrariGrassi}. Moreover, even if
this is the general situation in theories where there are no 
regularization  methods consistent with the symmetries, one would expect
to be possible to avoid this problem in QED and QCD.

\item A third approach was developed in a series of paper by Bonini,
D'Atta\-nasio and Marchesini \cite{BDM,BDM.QED,BDM.YM}.
Here the formalism of the Legendre transformed
cutoff effective action was developed in order to 
give a proof of renormalizability simpler and closer to the usual one of
quantum field theory. However the point of view about the symmetries
is essentially that of Becchi (even if generalized to $\mu-$momentum
prescriptions and directly extended to the $\Gamma(\Phi;\Lambda)$ functional 
in \cite{Datta}).

\item A fourth approach was implemented by Reuter and Wetterich in the
formalism of the effective average action \cite{scalarQED}. 
Here the idea is of adding
background gauge fields to the action in order to have explicit 
background gauge-invariance. However, this approach is quite cumbersome
in concrete computations and, moreover, its perturbative implementation
is not so efficient. In fact it is well known that a  perturbative
implementation (see for instance \cite{Grassi})
of the background field method requires fine tuning
of both Slavnov-Taylor identities and background Ward 
identities.

\item A fifth approach was introduced by Ellwanger \cite{Ellw} (see also
\cite{Datta}, where the relation between this approach and the Becchi's point
of view is clarified, and \cite{Pernici} for a careful analysis of the QED
case). In this
point of view, the attention is on the quantification of the 
gauge-breaking term,
which can be estimated by using some modified Slavnov-Taylor
identities. This approach is very appealing since the broken identities 
can be used to extract various non-trivial informations: for instance
the form of the chiral anomaly both in non-supersymmetric and in supersymmetric
chiral gauge theories \cite{B.Vian,BV.super}.
Nevertheless, an analytical study of the breaking term is very difficult 
in general.
 
\item Finally, there is a recent proposal of Morris 
\cite{Morris.gauge} based on a fully 
gauge-invariant formalism where the fundamental quantities are
Wilson loops and Wilson lines. 
The idea is of combining the numerical methods avalaible for the Exact
Renormalization Group  with the insights coming from the large $N_C$ 
expansion, where $N_C$ is the number of colors.
However the analysis is not simple and the comparison with the perturbative
results is difficult; moreover 
by construction this approach cannot say nothing about the Abelian
case which is our concern in the present work.

\end{enumerate}
In this paper we provide an explicitly Ward-identities-consistent 
formulation of the evolution equation for the case of
Abelian QED-like  gauge theories. This formulation, if extended to 
non-covariant gauges, is also suitable for the analysis of 
the non-Abelian case. This is left for a separate publication \cite{paper.II}.

\section{Tree Level Quantum Electrodynamics}

As a typical example of Abelian gauge theory we will consider the 
quantum electrodynamics (QED) with fields $\Phi=(A,\psi,\bar\psi)$. 
Our notations on metric and gamma matrices are as in \cite{IZ} 
and the covariant derivative is $D_\mu=\partial_\mu-ieA_\mu$.
The electron mass is denoted by $m$ and
the gauge fixing parameter by $\xi$ ;
in explicit computations we will use the Feynman gauge $\xi=1$.
Some useful abbreviations on integrals are
\formula{int.abbr}
{\int_x=\int d^4x,\quad\int_p=\int\frac{d^4p}{(2\pi)^4}.
}
For the Euclidean momenta we use the notations 
\formula{Euclidean}
{q_E=(i q_0,\vec q),\quad q_E^2\equiv\delta_{\mu\nu}\ q_E^\mu\ q_E^\nu=-q^2.}
If not otherwise specified, all the quantities 
should be intended in the Minkowski space.

The fundamental ingredients of our analysis are:
\begin{enumerate}

\item The functional operator
\formula{W}
{\W_f=-\int_x f(x)\left[\partial_\mu\dd{}{A_\mu}+ie\bar\psi\dd{}{\bar\psi}-
ie\psi\dd{}{\psi}\right]}
which defines the gauge symmetry,
\formula{gauge.tr}
{\W_f A_\mu=\partial_\mu f,\quad \W_f \psi=ie f\psi,\quad
\W_f\bar\psi=-ief\bar\psi.}

\item  The classical gauge-invariant action 
\formula{S.CL}
{S_{CL}(\Phi)
=\int_x-\frac14 F_{\mu\nu}F^{\mu\nu}+\bar\psi i(\slash D-m)\psi,\quad\W_f
S_{CL}=0}
which specifies the theory. 

\item The infrared cutoff functions $\kIR(q)$ (for the photon propagator)
and $\tilde\kIR(q)$ (for the electron propagator)
which specify the distinction between soft ($q_E^2<<\Lambda^2$)
and hard ($q_E^2>>\Lambda^2$) modes. 
In general they are smooth functions
with the properties
\formula{cutoff1}
{\lim_{\Lambda\to\infty} \kIR(q)=0,\quad 
\lim_{\Lambda\to0} \kIR(q)=1,
}
\formula{cutoff2}
{\lim_{\Lambda\to\infty} \tilde\kIR(q)=0,\quad 
\lim_{\Lambda\to0} \tilde\kIR(q)=1,
}
i.e. soft momenta  are dumped whereas hard momenta 
are unaffected.
\end{enumerate} 

A comment about the cutoff functions is in order here.
Usually one does not distinguish between the 
cutoff function for photons and electrons, 
i.e. keeps $\kIR(q)=\tilde\kIR(q)$. 
Nevertheless this distinction will be important in our analysis.
In particular $\kIR(q)$ is a scalar function whereas
$\tilde\kIR(q)$ is intended as a matrix in spinor space. One could also take
$\kIR(q)$ as a matrix in Lorentz indices, but this generalization is not
needed here.
From the cutoff functions one derives the following useful 
objects
\formula{Q.vector}
{Q^{\mu\nu}_\Lambda(p)=-[\kIR^{-1}(p^2)-1]p^2g^{\mu\nu}-\frac1\xi p^\mu p^\nu
}
and
\formula{Q.spinor}
{\tilde Q_\Lambda(p)=[\tilde\kIR^{-1}(p)-1](\slash p-m).
}
They enter in the gauge-fixed tree level cutoff action as follows:
\formula{tree.level}
{\Gamma^{(0)}(\Phi;\Lambda)=S_{CL}(\Phi)+\frac12\tilde\Phi Q_\Lambda\Phi,
}
\formula{def.Q}
{\frac12\tilde\Phi Q_\Lambda\Phi\equiv
\frac12\int_p A_\mu(-p)Q^{\mu\nu}_\Lambda(p)A_\nu(p)+\int_p\bar\psi(p)
\tilde Q_\Lambda(p)\psi(p).
}
In general there is an hard gauge-invariance problem because of the
quadratic breaking term  $\frac12\tilde\Phi Q_\Lambda(p)\Phi$ 
(we remind that the $Q_\Lambda(p)$'s in
$x-$space are in general complicate non-local differential operators):
\formulona{Ward.broken}
{\W_f\Gamma^{(0)}&=\int_x A_\mu Q^{\mu\nu}_\Lambda(i\partial)\partial_\nu f+
\partial_\mu fQ^{\mu\nu}_\Lambda(i\partial)A_\nu\\&+\int_x
ie\bar\psi Q_\Lambda(i\partial)(f\psi)-ie\bar\psi fQ_\Lambda(i\partial)\psi.
}
The main idea of this paper is of solving this problem by using a
particularly simple form for $Q^{\mu\nu}_\Lambda(p)$ and 
$\tilde Q_{\Lambda}(p)$. In particular we will take
\formule{Q.simple}
{Q_\Lambda^{\mu\nu}(p)&=&\Lambda^2 g^{\mu\nu}-\frac1\xi p^\mu p^\nu,\\
\tilde Q_\Lambda&=&-i\Lambda\gamma_5,
}
which corresponds to the following choice for the cutoff 
functions in Minkowski space,
\begin{subequations}
\formula{k.ph}
{\kIR(p)=\frac{p^2}{p^2-\Lambda^2}}
\formula{k.fe}
{\tilde\kIR(p)=\frac{\slash p-m}{\slash p-m-i\Lambda\gamma_5}.}
\end{subequations}
With these choices the explicit expressions for the tree level propagators 
of the theory are 
\formulona{prop.fot}
{&D_{\mu\nu}(k;\Lambda)=\Gamma^{(0)}_{\mu\nu}(k;\Lambda)^{-1}=
\left((-k^2+\Lambda^2)g_{\mu\nu}+(1-\frac1\xi)k_\mu k_\nu\right)^{-1}\\&=-
\frac{g_{\mu\nu}}{k^2-\Lambda^2+ i\ep}+(1-\xi)\frac{k_\mu k_\nu}
{(k^2-\Lambda^2+i\ep)(k^2-\xi\Lambda^2+i\ep)}
}  
and
\formula{prop.elet} 
{S_{\alpha\beta}(p;\Lambda)=\Gamma^{(0)}_{\alpha\beta}(p;\Lambda)^{-1}=
\left(\slash p -m-i\Lambda\gamma_5\right)^{-1}
=\frac{(\slash{p}+m-i\Lambda\gamma_5)_{\alpha\beta}}{p^2-m^2-\Lambda^2
+ i\ep},
}
where the causal $i\ep$ prescription has been used.
In section \ref{Pauli-Villars} we will discuss the technical 
advantages of the choice (\ref{Q.simple}b) 
for the spinor ``mass'' term, involving
the $\gamma_5$ matrix. Essentially this choice will simplify the task of
finding a suitable ultraviolet regularization of the evolution equation.
However, notice that we cannot give a direct physical meaning to the effective 
action for $\Lambda\neq0$ since parity symmetry is broken (
the transformation law of the effective action under parity is 
$\Gamma(A',\psi',\bar\psi';\Lambda)=
\Gamma(A,\psi,\bar\psi;-\Lambda)$ ).

The important point we want to stress here, is the fact
that with this choice of the infrared
cutoff functions the breaking of gauge invariance is innocuous, 
in the sense that the gauge breaking term is {\it linear} in the fields:
\formula{Ward.tree}
{\W_f\Gamma^{(0)}=\int_x\left( \frac\square\xi+\Lambda^2\right)
\partial_\mu A^\mu f(x)
}
As it is well known \cite{ZJ.book}
this exceptional property guarantees Ward identities can be
lifted to the quantum level to all orders in perturbation theory. More in
detail, if we expand the effective action in powers or $\hbar$
\formula{pert.exp}
{\Gamma^{[\ell]}=\Gamma^{(0)}+\hbar\Gamma^{(1)}+\dots+\hbar^\ell
\Gamma^{(\ell)}
}
and we use a consistent renormalization procedure, we can have in general
\formula{Ward.gen}
{\W_f\Gamma^{[\ell]}=\int_x\left( \frac\square\xi+\Lambda^2\right)
\partial_\mu A^\mu f(x)\quad\forall\ \ell\geq0.
}
In other words, the breaking of gauge invariance is confined at the
tree level and the perturbative corrections to $\Gamma^{(0)}(\Phi;\Lambda)$ 
are gauge invariant: 
\formula{Ward.corr}
{\W_f\Gamma^{(\ell)}=0\quad\forall\ \ell\geq1;
}
This fact can be proved with the standard techniques
of perturbative quantum field theory \cite{Piguet}; however here we will give
a Wilsonian analysis based on the evolution equation.

\section{The evolution equation}

The Wilson's evolution equation has a long history 
\cite{Wilson,Wegner-H,Polchinski,Hasenfratz} and there are many 
different formulations which describe the same physics. 
Here we are interested in the Legendre transformed
version of the equation introduced in \cite{BDM}. A detailed explanation 
of the employed notations and some comments about the derivation 
are collected in appendix A. Here we report only the final form for
the proper vertices, which reads
\formula{eq.ev.ver}
{\dot\Pi_{A_1\dots A_n}=I_{A_1\dots A_n}=-\frac i2(-)^A
(\Gamma_2^{-1}\dot Q_\Lambda\Gamma_2^{-1})^{BA}
\bar\Gamma_{AA_1\dots A_nB}
}
where the dot denotes the $\LdL$ derivative and
\formula{def.Pi}
{\Pi(\Phi;\Lambda)\equiv\Gamma(\Phi;\Lambda)-\frac12(-)^A\Phi^A 
Q_{\Lambda,AB}\Phi^B.
}
The graphical interpretation is shown in figures 1 and 2.
\begin{figure}[t]                              
\setlength{\unitlength}{1.25 ex}
\begin{center}
\begin{picture}(44,8)

 \put(3,5){\framebox(4,2)}
 \puntino{1,6}{A}\put(1,6){\line(1,0){2}}
 \put(9,6){\line(-1,0){2}}\puntino{9,6}{B}
 \put(3,2){\line(1,3){1}}\puntino{3,2}{A_1}
 \puntino{7,2}{A_n}\put(7,2){\line(-1,3){1}}

 \scrivi{11,6}{=} \put(17,6){\oval(4,2)}
 \puntino{13,6}{A}\put(13,6){\line(1,0){2}}
 \put(21,6){\line(-1,0){2}}\puntino{21,6}{B}
 \put(15,2){\line(1,3){1}}\puntino{15,2}{A_1}
 \puntino{19,2}{A_n}\put(19,2){\line(-1,3){1}}

 \scrivi{23.5,6}{-} \put(30,6){\oval(4,2)}
 \puntino{26,6}{A}\put(26,6){\line(1,0){2}}
 \put(34,6){\line(-1,0){2}}\puntino{34,6}{CD}
 \put(28,2){\line(1,3){1}}\puntino{28,2}{A_{i_1}}
 \puntino{32,2}{A_{i_k}}\put(32,2){\line(-1,3){1}}

 \put(36,5){\framebox(4,2)}
 \put(42,6){\line(-1,0){2}}\puntino{42,6}{B}
 \put(34,6){\line(1,0){2}}\puntino{34,6}{CD}
 \put(36,2){\line(1,3){1}}\puntino{36,2}{A_{i_{k+1}}}
 \puntino{40,2}{A_n}\put(40,2){\line(-1,3){1}}
 
\end{picture} \caption{{\small Recursive expansion of the
$\bar\Gamma_{AA_1\dots A_nB}$ vertices, denoted by the boxes. 
The black dots denote the full propagators and the ovals the full vertices.}}
\end{center}
\end{figure}
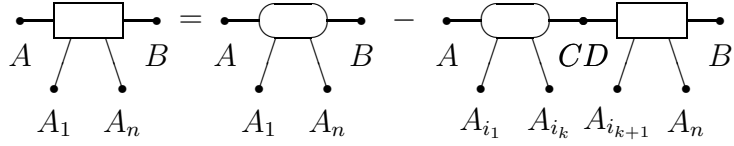

\begin{figure}[t]                              
\setlength{\unitlength}{1.2 ex}
\begin{center}
\begin{picture}(40,12)

 \scrivi{-1,6}{\frac 2i\LdL}
 \put(5,6){\oval(4,2)}\scrivi{5,6}{}
 \put(3,2){\line(1,3){1}}\puntino{3,2}{A_1}
 \puntino{7,2}{A_n}\put(7,2){\line(-1,3){1}}

 \put(17,6){\oval(8,6)[t]} \scrivi{17,9}{X}
 \scrivi{10,6}{=} \put(17,6){\oval(4,2)}\scrivi{17,6}{}
 \puntino{13,6}{}\put(13,6){\line(1,0){2}}
 \put(21,6){\line(-1,0){2}}\puntino{21,6}{}
 \put(15,2){\line(1,3){1}}\puntino{15,2}{A_1}
 \puntino{19,2}{A_n}\put(19,2){\line(-1,3){1}}

 \put(34,6){\oval(16,10)[t]} \scrivi{34,11}{X}
 \scrivi{23.5,6}{-} \put(30,6){\oval(4,2)}\scrivi{30,6}{}
 \puntino{26,6}{}\put(26,6){\line(1,0){2}}
 \put(34,6){\line(-1,0){2}}\puntino{34,6}{}
 \put(28,2){\line(1,3){1}}\puntino{28,2}{A_{i_1}}
 \puntino{32,2}{A_{i_k}}\put(32,2){\line(-1,3){1}}
 \put(36,5){\framebox(4,2)}
 \put(42,6){\line(-1,0){2}}\puntino{42,6}{}
 \put(34,6){\line(1,0){2}}\puntino{34,6}{}
 \put(36,2){\line(1,3){1}}\puntino{36,2}{A_{i_{k+1}}}
 \puntino{40,2}{A_{i_n}}\put(40,2){\line(-1,3){1}}
\end{picture} 
\end{center} \caption{{\small
Diagrammatic version of the exact 
evolution equation in Minkowski space. Here $X=\dot Q_\Lambda$.}}
\end{figure}
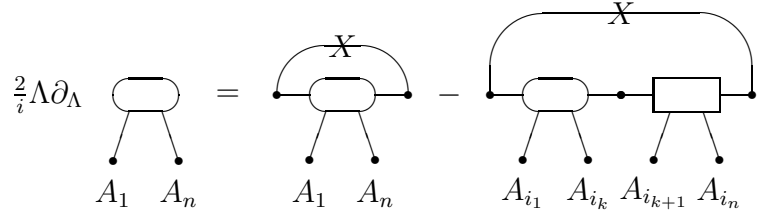

The generalized matrix $Q_{\Lambda,AB}$ which enters
in the evolution equation \rif{eq.ev.ver}
can be at large extent generic, depending 
on the cutoff functions choice via equations \rif{Q.vector} and \rif{Q.spinor}.
In order to preserve gauge-invariance we would like to use the 
Ward-identities-consistent mass cutoff \rif{Q.simple}. However, 
it can be seen by direct inspection that this choice does not sufficiently
regularize the ultraviolet behavior of the loop integral in two-point
functions.
Therefore we will invoke, as an intermediate step, an ultraviolet
regularization such as momentum integrals are well defined. Since in
the case we are studying ultraviolet regularizations consistent with
gauge symmetry do exist (in the following 
we will introduce a Pauli-Villars-like
regularization), this step gives no problems. 
The renormalizability property guarantees that the intermediate
regularization can be removed at the end, once the correct subtractions
are performed. 

We notice also that in theories with better ultraviolet behavior, such
as low dimensional or supersymmetric theories, 
this intermediate step can be skipped.

\section{Boundary conditions and perturbative expansion}

Now we review in brief how the evolution equation \rif{eq.ev.ver}
can be solved iteratively, once having specified suitable
boundary conditions. Since the fixing of the boundary condition is 
a non-trivial point, we report here some technical remarks 
(see also \cite{BDM} and
the original discussion of Polchinski \cite{Polchinski})~.

\begin{enumerate}
\item We split the effective action in a relevant and an irrelevant
part
\formulaX
{\Gamma(\Phi;\Lambda)=\Gamma_{rel}(\Phi;\Lambda)+\Gamma_{irr}(\Phi;\Lambda)
}
where by definition the relevant part contains only renormalizable 
interactions, i.e. terms with couplings of non-negative mass-dimension,
\formula{Pi.rel}
{\Gamma_{rel}(\Phi;\Lambda)=\sum_r c_r(\Lambda)\int_xO_r[\Phi],\quad\dim 
c_r(\Lambda)\geq0,
} 
where $O_r[\Phi]$ denote the (finite) set of  
relevant local operators 
($\dim O_r[\Phi]\\ \leq4$) constructed with the fields and their derivatives
which are consistent with the symmetries. 
In particular we extract the relevant part 
by using zero-momentum prescriptions, i.e.
by a Taylor expansion in fields and momenta (see appendix B for details).
In this way $\Gamma_{irr}(\Phi;\Lambda)$ 
contains only couplings with negative mass dimension. However other
prescriptions are possible, as for instance on-shell renormalization 
prescriptions \cite{BDM.QED} or prescriptions at momentum $\mu$ \cite{BDM.YM}.

\item Following Polchinski, we suppose 
of knowing the relevant part of the action, i.e. the relevant
parameters, at some initial low-energy scale $\Lambda_R$,
which can be thought as the typical energy scale accessible in every day
experiments\footnote{As a matter or fact, it is also possible to fix 
the couplings at the scale $\Lambda=0$. But in this case and in presence of 
massless particles, one is forced to introduce a non-zero momentum scale
$\mu$ as a subtraction point. That scale plays the same role of $\Lambda_R$.}.
On the other hand, the irrelevant parameters 
are fixed at some ultraviolet scale
$\Lambda_{UV}>>\Lambda_R$, which is interpreted as 
the scale where new physics (unification, quantum gravity, etc) 
is expected to modify completely our field theory. 
By dimensional arguments one expects that the
irrelevant couplings affect the low-energy Green functions only as
inverse powers of $\Lambda_{UV}$ and in fact this can be rigorously
proved to all orders in perturbation theory, as done for the first
time by Polchinski. Therefore we can safely take
\formulaX
{\Gamma_{irr}(\Phi;\Lambda_{UV})=0
}
for large $\Lambda_{UV}$.
\end{enumerate}
Having stated the boundary conditions we can
write the exact evolution equation in its integral form \cite{BDM}
\formule{eq.int.rel}
{\Gamma_{rel}(\Phi;\Lambda)&=&\left.\Gamma_{rel}\right|_{\Lambda=\Lambda_R}
+\int_{\Lambda_R}^{\Lambda}\frac{d\Lambda_1}{\Lambda_1} 
I_{rel}(\Phi;\Lambda_1)\\ \label{eq.int.irr}
\Gamma_{irr}(\Phi;\Lambda)&=&\left.\Gamma_{irr}\right|_{\Lambda=\Lambda_{UV}}
-\int_{\Lambda}^{\Lambda_{UV}}\frac{d\Lambda_1}
{\Lambda_1}I_{irr}(\Phi;\Lambda_1).
}
Now we are in position to solve iteratively the  ERGE
by expanding the effective action in the loop-wise
series \rif{pert.exp}.
In this way the integrated evolution equation \rif{eq.int.rel}, 
with the following
boundary conditions on relevant and irrelevant couplings 
\formula{b.c}
{\left.c_r^{(\ell)}\right|_{\Lambda=\Lambda_R}=\bar c_r\delta_{\ell0},\quad 
\left.c_i^{(\ell)}\right|_{\Lambda=\Lambda_{UV}}\equiv0,
}
can be solved iteratively:
\formula{eq.int.2n}
{\Gamma^{(\ell+1)}(\Phi;\Lambda)=\int_{\Lambda_R}^\Lambda
\frac{d\Lambda_\ell}{\Lambda_\ell}I^{(\ell)}_{rel}
-\int_\Lambda^{\Lambda_{UV}}
\frac{d\Lambda_\ell}{\Lambda_\ell}I^{(\ell)}_{irr},\quad\ell\geq0.
} 
In particular, as we will prove in section 8, in the perturbative expansion
one can safely replace $\Lambda_{UV}=\infty$ and the dependence on the
ultraviolet scale is completely lost.\\

\section{The $\phi^4_4$ theory}

In order to see how the previous general analysis works in a simple example,
we consider here 
the paradigmatic case of the Euclidean massless $\phi^4_4$ theory,
regularized in the infrared with a mass-like cutoff $\Lambda^2$ and
in the ultraviolet through an higher derivative regularization. 
In addition, we introduce
an external source $K(x)$ coupled to the composite operator $\phi^2(x)/2$.

We take as tree-level Euclidean action of the model
\formulona{action}
{\Gamma^{(0)}(\phi,K;\Lambda,M_0)&=
\int_x\frac12\partial^\mu\phi\partial_\mu\phi+
\frac\lambda{4!}\phi^4+
\frac12\phi\frac{(\partial^2)^2}{M_0^2}\phi\\ &\qquad
+\frac12\Lambda^2\phi^2+\frac12K(x)\phi^2.
}
The regularized free propagator reads
\formula{HD.prop}
{D_{reg}(q;\Lambda,M_0)=\frac1{q^2+\Lambda^2+q^4/M_0^2}
}
and satisfies the important property 
\formula{dotD}
{\dot D_{reg}(q;\Lambda,M_0)\simeq -2\Lambda^2 \frac{M_0^4}{q^8},
\quad q^2>>M_0^2.
}
This fact is essential to verify that for any finite $M_0$ 
all the momentum integrals 
in the evolution equation are well defined in the ultraviolet. 

In the following the $M_0$-dependence will be often understood.
We will denote the proper vertices with $l$ insertions of the
operator $\phi^2/2$ as
\formula{G.n.l}
{\Gamma_{2n,l}(x_i,y_j;\Lambda)\equiv\left.\dd\Gamma
{\phi(x_1)\dots\delta\phi(x_{2n})
\delta K(y_1)\dots\delta K(y_l)}\right|_{\phi=K=0}
}
and their Fourier transforms as
\formula{Gamma.n.l}
{\tilde\Gamma_{2n,l}(p_i,q_j;\Lambda)=(2\pi)^4
\delta\left(\sum_i p_i+ \sum_j q_j\right)\Gamma_{2n,l}(p_i,q_j;\Lambda).
}
Notice that the derivative with respect to $\Lambda^2$ can be replaced
with a derivative with respect to $K$ taken at zero momentum; for instance
we have $\partial_{\Lambda^2}\Gamma^{(0)}_{2n}(p_i;\Lambda)=
\Gamma^{(0)}_{2n,1}(p_i,0;\Lambda)$. In general the relation 
\formula{der.K}
{\partial_{\Lambda^2}\Gamma_{2n,l}(p_i,q_j;\Lambda)=
\Gamma_{2n,l+1}(p_i,q_j,0;\Lambda)
}
can be imposed to all orders in perturbation theory.
We define the relevant coefficients
\formule{rel.c}
{c_m(\Lambda)=\Gamma_2|_{p=0},\quad\sigma_m(\Lambda)=\Gamma_{2,1}|_{p_i=0},\\
c_\phi(\Lambda)=\partial_{p^2}\Gamma_2|_{p=0},\quad
c_\lambda(\Lambda)=\Gamma_4|_{p_i=0},
}
the relevant action
\formula{G.rel}
{\Gamma_{rel}(\phi,K;\Lambda,M_0)=
\int_x\frac12c_\phi\partial^\mu\phi\partial_\mu\phi+
\frac1{4!}c_\lambda\phi^4+\frac12c_m\phi^2+\frac12\sigma_m K\phi^2
}
and the irrelevant vertices
\formule{irr.v}
{\Gamma_{2,irr}(p;\Lambda)&=&\Gamma_2(p;\Lambda)-
\Gamma_2|_{p=0}-p^2\partial_{\bar p^2}\Gamma_2|_{\bar p=0},\\
\Gamma_{4,irr}(p_i;\Lambda)&=&\Gamma_4(p_i;\Lambda)-\Gamma_4|_{p_i=0},\\
\Gamma_{2,1,irr}(p,q;\Lambda)&=&\Gamma_{2,1}(p,q;\Lambda)-
\Gamma_{2,1}|_{p=q=0},\\
\Gamma_{2n,l,irr}(p_i,q_j;\Lambda)&=&
\Gamma_{2n}(p_i,q_j;\Lambda),\quad n>2,\ l>1.
}
The boundary conditions on the relevant couplings (renormalization 
prescriptions) at loop $\ell$ are
\formule{ren.prescr}
{c_m^{(\ell)}(0)=&\Gamma_2^{(\ell)}|^{\Lambda=0}_{p=0}&=0,\\
\sigma^{(\ell)}_m(\Lambda_R)=&\Gamma^{(\ell)}_{2,1}
|^{\Lambda=\Lambda_R}_{p=0}&=\delta_{\ell0},\\
c^{(\ell)}_\phi(\Lambda_R)=&\partial_{p^2}\Gamma^{(\ell)}_2
|^{\Lambda=\Lambda_R}_{p^2=0}&=\delta_{\ell0},\\
c^{(\ell)}_\lambda(\Lambda_R)=&\Gamma^{(\ell)}_4
|^{\Lambda=\Lambda_R}_{p_i=0}&=\lambda\delta_{\ell0},
}
where the renormalization scale $\Lambda_R<<M_0$ is non-zero in order to
avoid infrared divergences.
The irrelevant couplings are fixed 
at the ultraviolet scale $\Lambda_{UV}$ which is of order $M_0$.

Now we make some specific one-loop computation.
At leading order approximation the evolution equation for the vertices
without insertions has the explicit form 
\formula{I.scal.0}
{\dot\Pi^{(1)}_{2n}(p_i;\Lambda)=-\int_q
\frac{\Lambda^2}{(\Lambda^2+q^2+q^4/M_0^2)^2}
\bar\Gamma_{2n+2}^{(0)}(q,p_i,-q;\Lambda)
}
where the recursive form of the $\bar\Gamma^{(0)}_{2+2n}$ vertices
is given by the condensed expression (see figure 1)
\formulona{bar.Gamma.n}
{\bar\Gamma^{(0)}_{2}&=0,\quad\bar\Gamma_4^{(0)}=\lambda,\\
\bar\Gamma^{(0)}_{2+2n}&=\Gamma^{(0)}_{2+2n}-\sum_{k=1}^{n-1}
\Gamma^{(0)}_{2+2k}(\Gamma_2^{(0)})^{-1}
\bar\Gamma^{(0)}_{2+2n-2k},\quad n\geq2.
}
In particular the one-loop two-point equation reads
\formula{der.tadpole}
{\dot\Pi_2^{(1)}(p;\Lambda;M_0)=\int_{q}\frac{-\lambda\Lambda^2}
{(q^2+\Lambda^2+q^4/M_0^2)^2,}
}
and is logarithmically divergent,
\formulaX
{\dot\Pi_2^{(1)}(p;\Lambda;M_0)=
-\frac{\lambda\Lambda^2}{16\pi^2}\log\frac{M_0^2}{\Lambda^2}+
O(1).
}
This divergence can be compensed if we
introduce a mass renormalization coupling 
$Z_m^{(1)}(M_0/\Lambda_R)$ by defining
\formula{Pi.Gamma}
{\Gamma_2^{(1)}(p;\Lambda)\equiv\Lambda^2Z_m^{(1)}(M_0/\Lambda_R)
+\Pi_2^{(1)}(p;\Lambda).}
In this way the two-point equation for $\Gamma_2(p;\Lambda)$ 
\formula{two.point.eq}
{\dot\Gamma_2^{(1)}(p;\Lambda)=2Z_m^{(1)}(M_0/\Lambda_R)\Lambda^2+
2\Lambda^2\partial_{\Lambda^2}\Pi_2^{(1)}(p;\Lambda)
}
can be made finite by using \rif{der.K} and 
imposing the normalization condition (\ref{ren.prescr}b),
\formula{mass.norm.1}
{\left.\Gamma_{2,1}^{(1)}\right|_{p=0}^{\Lambda=\Lambda_R}=\left.
\partial_{\Lambda^2}\Gamma_2^{(1)}\right|_{p=0}^{\Lambda=\Lambda_R}=
Z_m^{(1)}(M_0/\Lambda_R)+
\left.\partial_{\Lambda^2}\Pi_2^{(1)}\right|_{p=0}^{\Lambda=\Lambda_R}=
0.
}
In this way by using \rif{der.tadpole} we obtain
\formula{Z_m.1}
{Z_m^{(1)}(M_0/\Lambda_R)=
\frac12\int_q\frac{\lambda}{(q^2+\Lambda_R^2+q^4/M_0^2)^2}
}
and the two-point evolution equation 
\rif{two.point.eq} has the explicit finite form
\formulaX{
\lim_{M_0\to\infty}\dot\Gamma_2^{(1)}=-\lambda\int_{q}\left[
\frac{\Lambda^2}{(q^2+\Lambda^2)^2}-\frac{\Lambda^2}{(q^2+\Lambda_R^2)^2}
\right]=\frac{\lambda\Lambda^2}{16\pi^2}\ln\frac{\Lambda^2}{\Lambda_R^2}.
}
Notice that only after the imposition of the renormalization prescription
(\ref{ren.prescr}b) the ultraviolet regularization
can be removed: the situation here is more similar to the traditional approach
to Quantum Field Theory than to the Wilsonian one. The difference is that
in the usual Wilsonian
formulation the $\LdL$ derivative of the
cutoff function is strongly damped in the ultraviolet and 
\rif{der.tadpole}
is automatically finite; at a consequence  both $\Gamma(\phi;\Lambda)$ and $
\Pi(\phi;\Lambda)$ are finite and the introduction of the renormalization
constant $Z_m(M_0/\Lambda_R)$ is not needed. This simplifies for certain
aspects the analisys,
but the price to pay is the lost of gauge-invariance. For this reason the
mass cutoff should be preferred in gauge-theories.

The higher points vertices are automatically finite at one-loop; 
for instance the four point vertex reads
\formulona{four.vertex}
{\lim_{M_0\to\infty} \dot\Gamma_4^{(1)}(p_i;\Lambda,M_0)&=
-\frac{\lambda^2}2\sum_{i<j}\int_q \dot D(q;\Lambda) 
D(q+p_i+p_j;\Lambda)\\
&=-\frac{\lambda^2}{32\pi^2}\sum_{i<j}\int_0^1 dx
\frac{\Lambda^2}{(p_i+p_j)^2 x(1-x)+\Lambda^2}.
}
{\it A fortiori} the finiteness property holds for the
vertices with insertions.

The analysis of higher order correction is more involved. The general
form of the evolution equation with $Q_\Lambda=
Z_m(M_0/\Lambda_R)\Lambda^2$ is
\formula{I.scal}
{\dot\Pi_{2n}(p_i,K;\Lambda)=-\int_q
\frac{Z_m\Lambda^2}{[Z_m\Lambda^2+\Pi_2(q,K;\Lambda)]^2}
\bar\Gamma_{2n+2}(q,p_i,-q,K;\Lambda)
}
where the mass renormalization coupling
\formula{Z_m.pert}
{Z_m(M_0/\Lambda_R)=1+Z_m^{(1)}(M_0/\Lambda_R)+Z_m^{(2)}(M_0/\Lambda_R)+\dots
}
is fixed by the renormalization prescription (\ref{ren.prescr}b) i.e. by 
the self-consistent equation, to be solved in perturbation theory,
\formula{Z_m}
{Z_m(M_0/\Lambda_R)=1+
\frac12\int_q\frac{Z_m\Gamma_4(q,0,0,-q;\Lambda_R)}{[Z_m\Lambda_R^2+
\Pi_2(q,\Lambda_R)]^2}.
}
Apparently for $M_0\to\infty$  the evolution equation
\rif{I.scal} contains overlapping divergences at higher orders
in perturbation theory. Actually, thanks to the renormalizability
proof, they cancel. This can be shown for instance by using
the Callan's  proof \cite{Callan} presented in 
\cite{ZJ.book}. Actually our approach can be seen as the bridge between
the Wilsonian point of the view and the Field Theory methods based on
the Callan-Symanzik equation. We refer to section 9 for more details
on this point. We stress here that the relevance of the Wilsonian
interpretation is the fact that there are numerical techniques, based
on suitable truncations\footnote{There is a subtle point here. Even if
the evolution equation \rif{I.scal} is well defined to all orders 
in perturbation theory as $M_0\to\infty$, this property relies on delicate
cancellations of Feynman diagrams and could be lost by using a generic
non-perturbative truncation. Nevertheless we checked explicitly that in the
typical non-perturbative approximation used in the Wilsonian literature, i.e.
the local potential approximation, there are no practical problems in
renormalizing the potential, by using for instance the 
Coleman-Weinberg prescriptions.}, to solve
the exact equation \rif{I.scal} {\it non-perturbatively}.

\section{The QED case} \label{Pauli-Villars}

The previous considerations generalize quite straightforwardly to the QED 
case, provided that we use a gauge-consistent ultraviolet regularization.
Here we will use a kind of Pauli-Villars 
regularization that we shall call 
holomorphic Pauli-Villars regularization following \cite{Zakharov}.

In general, the Pauli-Villars approach
consists in adding some very massive ($M_0>>m$) unphysical fields 
to the physical theory, in such a way of smoothing its ultraviolet
behavior.
In the case we are considering it is sufficient to 
take as tree level action
\formulona{PV.action}
{&\Gamma^{(0)}(A,\psi,\bar\psi,A',\psi',\bar\psi';\Lambda,M_0)=\\
&\int_x-\frac14F_{\mu\nu}F^{\mu\nu}-\frac1{2\xi}(\partial_\mu A^\mu)^2+\frac12
\Lambda^2A_\mu A^{\mu}+\\
&\int_x\bar\psi i\slash D(A+A')\psi-\bar\psi(m+i\Lambda\gamma_5)\psi+\\
&\int_x\frac14F'_{\mu\nu}F^{'\mu\nu}+\frac1{2\xi}(\partial^\mu A_\mu')^2-
\frac12(\Lambda^2+M_0^2)A_\mu'A^{\mu'}+\\&
\int_x\bar\psi' i\slash D(A+A')
\psi'-\bar\psi'(M_0+i\Lambda\gamma_5)\psi'
}
where we have introduced an heavy photon field $A'_\mu$ (commuting) and 
two heavy fermion fields $\psi',\bar\psi'$ (commuting) interacting in 
a gauge-invariant way 
(the covariant derivative is $D_\mu(A+A')=\partial_\mu-ie A_\mu-ieA_\mu'$).
The kinetic term of the heavy photon has a minus sign
compared to the kinetic term of the physical photon, thus
the effect of the unphysical photon fields $A_\mu'$  is 
a modification of the photon propagator at high energies
\formula{PV.prop}
{\left[D_{\mu\nu}(k;\Lambda^2)\right]_{reg}=D_{\mu\nu}(k;\Lambda^2)-
D_{\mu\nu}(k;\Lambda^2+M_0^2)
}
which becomes more convergent ($\left[D_{\mu\nu}\right]_{reg}\sim M_0^2/k^4$). 
The unphysical fermions are taken to be commuting therefore contributing 
with a plus sign to the fermion loops. The net effect is of increasing 
the ultraviolet convergence of the fermion bubble of a factor 
$M^2/q^2$. Notice that the infrared cutoff is inserted by means of the 
$\gamma_5$ matrix in order to ensure this property. As an intriguing 
additional bonus one has some 
analyticity properties on the dependence on the complex masses 
$m+i\Lambda\gamma_5$ and $M_0+i\Lambda\gamma_5$. This is very appealing 
in view of a supersymmetric extension of this work. A similar 
observation can be found in \cite{Zakharov}.

With the regularization we have introduced all the momentum
integrals in the evolution equation becomes well defined. 
Notice that in a more usual context,
in order to directly regularize the Feynman diagrams, a much more complicate
Pauli-Villars regularization with more unphysical fields is needed. However,
for our aims it is sufficient to regularize the ultraviolet behavior of the 
evolution equation i.e. of the {\it mass-derived} Feynman diagrams, which 
are more convergent of the standard ones. This simplifies our task.

At the end the intermediate regularization can be removed, provided that
we correctly subtract the divergences: this is done by imposing the 
zero-momentum renormalization prescriptions.
\begin{figure}[t]  
  \begin{center}\begin{tabular}{ccc}
  \mbox{\epsfig{file=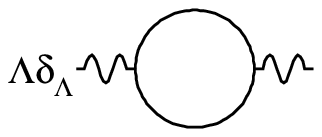,height=2.2cm}}&
  \mbox{\epsfig{file=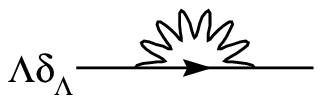,height=2.5cm}} &
  \mbox{\epsfig{file=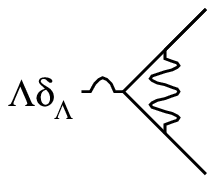,height=3cm}}
  \end{tabular} \end{center}
\caption{{\small 
Graphic representation of the right hand side of equation \rif{eq.evol.QED}. 
Each diagram should be subtracted by an analogous diagram involving the
Pauli-Villars heavy fields.
}}
\end{figure}

It is very simple to write down
the explicit form of the QED evolution equations
\rif{eq.ev.ver}
in the one-loop approximation. In fact by using the property\footnote{We 
remind that equation
\rif{prop} is a condensed matrix relation and means that analogous
relations holds both for the photon and the electron propagator.}
\formula{prop}
{-\frac1{\Gamma_2^{(0)}}\dot Q_\Lambda\frac1{\Gamma_2^{(0)}}=
\LdL\frac1{\Gamma_2^{(0)}}
}
one easily realize that the
right hand side of the one loop evolution equation 
is given by the logarithmic derivative of
the usual Feynman diagrams (minus a zero-momentum subtraction for the
two-point functions). Graphically the situation is represented in
figure 3. 
Since a detailed derivation can be found in 
\cite{BDM.QED} here we just report the formulae for the simpler proper
vertices:
\formulona{eq.evol.QED}
{\!\!\!\!\!\!\!\!\!
I_{\mu\nu}^{(1)}=& ie^2 \int_q \LdL
\Tr[\gamma_\mu S(q;\Lambda)\gamma_\nu S(q+p;\Lambda)]_{PV} \\
\!\!\!\!\!\!\!\!\!
I_{\alpha\beta}^{(1)}=& 
ie^2\int_q\LdL \left[D^{\mu\nu}(q;\Lambda)\gamma_\nu S(q+p;\Lambda)
\gamma_\mu e^2\right]_{\alpha\beta,PV}
\\\!\!\!\!\!\!\!\!\!
I_\mu^{(1)}=& ie^3\int_q\LdL\left[\gamma_\nu
D^{\nu\rho}(q;\Lambda)\gamma_\mu S(q+p;\Lambda)\gamma_\rho S(q+p';\Lambda)
\right]_{PV}.
}
The foot $PV$ remind that we are using the Pauli-Villars
regularization to properly define the two-point vertices.
As a concrete example, we can compute in detail
the evolution of the (inverse) photon propagator.
That analysis should be intended also as a practical introduction about 
checking Ward identities and computing beta functions in this
formalism.

The photon propagator can be decomposed in its transversal and 
longitudinal components (for details on notation see appendix B)
\formula{I_L.I_T}
{\Pi_T(p;\Lambda)=\frac13 t^{\mu\nu}(p)\Pi_{\mu\nu}(p;\Lambda),\quad
\Pi_L(p;\Lambda)=\ell^{\mu\nu}(p)\Pi_{\mu\nu}(p;\Lambda).
}
Doing the traces, using Feynman parameterization, and continuing to the 
Euclidean space, we can explicitly compute $\dot\Pi_T(p_E;\Lambda)$
and $\dot\Pi_L(p_E;\Lambda)$. 
In particular we obtain
\formula{I_T1}
{\dot\Pi_T^{(1)}(p_E;\Lambda)=
\int_0^1dx\int_{q_E}\LdL\left[ f_T(q_E,p_E;\Lambda,m)-
f_T(q_E,p_E;\Lambda,M_0)\right]
}
and
\formula{I_L1}
{\dot\Pi_L^{(1)}(p_E;\Lambda)=
\int_0^1dx\int_{q_E}\LdL\left[ f_L(q_E,p_E;\Lambda,m)-
f_L(q_E,p_E;\Lambda,M_0)\right]
}
where
\formulaX
{f_T(q_E,p_E;\Lambda,M)=4e^2\frac{\frac12q_E^2-p_E^2 x(1-x)+
M^2+\Lambda^2}{(q_E^2+p_E^2\ x(1-x)+M^2+\Lambda^2)^2}
}
and
\formulaX
{f_L(q_E,p_E;\Lambda,M)=4e^2
\frac{\frac12q_E^2+p_E^2 x(1-x)+M^2+\Lambda^2}
{(q_E^2+p_E^2\ x(1-x)+M^2+\Lambda^2)^2}.
}
Notice that, after the subtraction and the application of the $\LdL$ 
operator, the integrals in $q_E$ are perfectly convergent, 
and simple to compute. The final results are 
\formula{I_T}
{\dot\Pi_T^{(1)}(p_E;\Lambda)=-\frac{e^2}{\pi^2}\int_0^1dx
\frac{p_E^2\ x(1-x)\ \Lambda^2}{p^2_E\ x(1-x)+m^2+\Lambda^2}+O(1/M_0^2)
}
and
\formula{I_L}
{\dot\Pi_L^{(1)}(p_E;\Lambda)\equiv0,\quad\forall\ \Lambda,\ 
\forall\ p_E,\ \forall\ M_0<\infty.
}
The first result is remarkable because of the relation with the usual
renormalization group. In fact, for large $\Lambda$, the coefficient
in front to $I_T(p;\Lambda)$ is related to the one-loop QED beta 
function
\formulaX
{I_T(p_E^2;\Lambda^2)= 
-\frac83\frac{e^2}{16\pi^2}p_E^2+
O\left(\frac{p_E^2}{\Lambda^2},\frac{m^2}{\Lambda^2}\right).
}
The second result is also remarkable, because it is a direct check
of gauge-invariance, i.e. of the Ward identity
$$
p^\mu I_{\mu\nu}^{(1)}(p;\Lambda)\equiv0.
$$ 
Technically equation \rif{I_L} holds since the function 
$f_L(q_E,p_E;\Lambda,m)-f_L(q_E,p_E;\\ \Lambda,M_0)$ can be
rewritten as a total derivative,
\formula{tot.der}
{\dede{}{q^\mu}\left[\frac{q^\mu}{q^2+p_E^2 x(1-x)+m^2+\Lambda^2}-
\frac{q^\mu}{q^2+p_E^2 x(1-x)+M_0^2+\Lambda^2}\right]
}
and therefore its momentum integral is identically zero.
Here one sees the importance of the intermediate ultraviolet regularization:
had we not taken in account the Pauli-Villars fields, i.e. had we
neglected the second piece in \rif{tot.der}, should we have obtained a 
finite but wrong (non-zero) result.
Similar subtleties are well known in the literature. The same
remark on the necessity of regularizing the evolution equation, even
if in a very different formalism, can be found in \cite{Morris.gauge}.

Having explained how the machinery works on simple examples, now we can turn
to the analysis of general questions, like gauge-invariance and 
renormalizability.

\section{The gauge-invariance proof}

It is quite simple to prove that our formulation is  consistent with
the gauge symmetry, i.e. that the $\Gamma(\Phi;\Lambda)$ functional 
is gauge invariant
\formula{Ward.QED}
{\W_f\Gamma(\Phi;\Lambda)=0\quad\forall\Lambda
} 
for {\it any} $\Lambda$. This fact is expected, since the perturbative
expansion of the evolution equation gives the usual Feynman
diagrams with massive propagators
supplemented with the BPHZ zero-momentum subtractions and it 
is known that this approach is consistent with Ward-Takahashi identities
\cite{Lowenstein.QED}. However here we will give a more direct proof 
based on the evolution equation.\\
The simpler way to proceed is from diagrammatic considerations, even if more
formal non-diagrammatic proofs are possible \cite{paper.II}. 

At the level of proper vertices the functional Ward
identity \rif{Ward.QED} corresponds to an infinite set of 
transversality constraints like
\formulona{trans}
{&k^\mu\Pi_{\mu\nu}(k;\Lambda)=0,\quad (k_1+k_2+k_3)^\mu 
\Gamma_{\mu\nu\rho\lambda}
(k_1,k_2,k_3;\Lambda)=0,\\
&(p_2-p_1)^\mu\Gamma_{\mu\alpha\beta}(p_1,-p_2;\Lambda)=
e\Gamma_{\alpha\beta}(p_2;\Lambda)-
e\Gamma_{\alpha\beta}(p_1;\Lambda),
\\&(k+p_1+p_2)^\mu\Gamma_{\mu\nu\alpha\beta}(k,p_1,p_2;\Lambda)=
e\Gamma_{\nu\alpha\beta}(k,p_1;\Lambda)-
e\Gamma_{\nu\alpha\beta}(k,p_2;\Lambda),
}  
and infinite others. In general the transverse part 
of a vertex $\Gamma_{n_A+1,n_{\bar\psi\psi}}(k_1\\ \dots k_{n_A+1},p_1\dots 
p_{2n_{\bar\psi\psi}})$ 
is related to a difference of vertices 
$\Gamma_{n_A,n_{\bar\psi\psi}}(k_1\dots k_{n_A},\\ p_1\dots 
p_{2n_{\bar\psi\psi}})$ or, in absence of fermion legs, is zero. 

It is clear that a proof of explicit gauge-invariance 
is doomed to fail for 
a {\it generic} choice of cutoff functions. In fact in the generic case
the Ward identities are badly broken {\it at tree level}; for instance
the vertex Ward identity does not hold
\formulona{Ward30.gen}
{\Gamma^{(0)}_{\alpha\beta}(p_1;\Lambda)-\Gamma^{(0)}_{\alpha\beta}
(p_2;\Lambda)&=\kIR^{-1}(p_1)(\slash p_1-m)-\\
&\quad\kIR^{-1}(p_2)(\slash p_2-m)\\
&\neq(p_1^\mu-p_2^\mu)\Gamma^{(0)}_\mu(p_1,p_2;\Lambda)
}
therefore there is no hope to recover gauge-invariance at {\it any} $\Lambda$.
On the other hand, with our choice of cutoff 
functions, the situation is much better and 
for instance the tree level vertex satisfies the correct transversality 
relation
\formulonaX
{\Gamma^{(0)}_{\alpha\beta}(p_1;\Lambda)-\Gamma^{(0)}_{\alpha\beta}
(p_2;\Lambda)&=
\left[\slash p_1-m-i\Lambda\gamma_5\right]-\left[\slash p_2-m-
i\Lambda\gamma_5\right]\\
&=(p_1^\mu-p_2^\mu)\Gamma^{(0)}_\mu(p_1,p_2;\Lambda)/e
}
for any $\Lambda$. This is obvious because 
a change of the fermion mass matrix from $m\to m+i\Lambda\gamma_5$ 
does not break gauge-invariance. With our
choice the only breaking of gauge-invariance is focused on the 
vertex $\Gamma_{\mu\nu}^{(0)}(k;\Lambda)$ which is not transverse,
\formula{tr.ph}
{k^\mu\Gamma_{\mu\nu}^{(0)}(k;\Lambda)=\left(-\frac {k^2}\xi+\Lambda^2\right)
k_\nu\neq0.}
However the perturbative corrections $\Gamma_{\mu\nu}^{(\ell)}$ are transverse.
Therefore we expect gauge-invariance be preserved for 
$\Gamma^{(\ell)}(\Phi;\Lambda),\ l\geq1$.

The logic of a formal proof is the following.
\begin{enumerate} 
\item We suppose that
the functional $\Gamma(\Phi;\bar\Lambda)$  is gauge-invariant (i.e. the
proper vertices satisfy Ward identities) at some initial scale $\bar\Lambda$. 
\item We observe that in this
hypothesis even the functional $I(\Phi;\bar\Lambda)$ is gauge invariant 
(i.e. the
$I_n(p_i;\bar\Lambda)$ vertices satisfy Ward identities) 
at the scale $\bar\Lambda$. 
\item Therefore
the evolution equation is gauge-invariant and, as a consequence, 
the Ward identities are satisfied to any $\Lambda$.
\end{enumerate}
One can convince himself of the transversality property of the
$I_n(p_i;\bar\Lambda)$ vertices
directly from their definition,
by considering some specific case like $I_{\mu\nu},\ I_{\mu\alpha\beta},$
etc. and by using the Ward identities \rif{trans} at the scale
$\bar\Lambda$.
For instance one can prove the transversality relation 
$k^\mu I_{\mu\nu}(k;\bar\Lambda)=0$. 
To this aim one must take in account {\it all} the pieces
in $I_{\mu\nu}(k;\bar\Lambda)$; moreover usual
tricks, such as the use of the cyclic property of the trace 
and the possibility of
doing translations in the momentum integrals must be applied.
Notice that 
this latter translation can readily be  done thanks to the intermediate 
regularization making convergent the integration.

The ultimate reason 
for the validity of all Ward identities is of geometric origin 
and it is completely elucidated in \cite{paper.II}. 

Notice that this proof does not require loop expansion and  formally holds
even non-perturbatively provided that all the momentum integrals
implicit in the evolution equation are well defined. This is guaranteed by 
the ultraviolet regularization. The possibility of removing the
regularization can be rigorously 
proved to all orders in perturbation theory, as shown in appendix C. 

\section{The Callan-Symanzik equation}

In this section we explain the relation between our formulation of the 
Wilson renormalization
group equation and the Callan-Symanzik equation \cite{CS}. 
Such a relation is expected because
the two approaches are very similar: in both case we study the response 
of the field theory (i.e. of the functional $\Gamma(\Phi,K;\Lambda)$) 
under variations of a mass term.

In order to simplify the notation, 
initially we consider the Euclidean massless $\phi_4^4$ 
theory and then we extend to the QED case.  
The first step to convert the ERGE in a form suitable for comparison with
the standard Callan-Symanzik equation consists in the introduction of
the rescaling functions (at zero-momentum)
\formula{Z}
{\hat Z_\phi(\Lambda)=\partial_{p^2}\Gamma_2|_{p=0},\quad
\hat Z_K(\Lambda)=\hat Z_\phi^{-1}(\Lambda)\Gamma_{2,1}|_{p_i=0},}
and of the flowing coupling (at zero-momentum)
\formula{coupling}
{\hat\lambda(\Lambda)=\hat Z_\phi^{-2}(\Lambda)\Gamma_4|_{p=0}.}
Now we define the rescaled quantities
\formula{riscalamento}
{\hat\phi=\hat Z_\phi^{1/2}(\Lambda)\phi,\quad\hat K=Z_K(\Lambda) K,\quad 
\hat Q_\Lambda=\hat Z_\phi(\Lambda)^{-1} Q_\Lambda.
}
With these redefinitions the relation 
$\hat\Gamma(\hat\phi,\hat K,\hat\lambda;\Lambda)=
\Gamma(\phi,K;\Lambda)$, i.e.
\formula{hat.rel}{
\hat\Pi(\hat\phi,\hat K,\hat\lambda;\Lambda)+
\frac12\hat\phi\cdot\hat Q_\Lambda\hat\phi=\Pi
(\phi,K;\Lambda)+\frac12\phi\cdot Q_\Lambda\phi,
} 
holds, therefore the proper vertices rescale as 
\formulaX
{\hat\Gamma_{2n,l}=\hat Z_\phi^{-n}Z_k^{-l}\Gamma_{2n,l}\; ,
\quad\hat{\bar\Gamma}_{2+2n,l}=\hat Z_\phi^{-n-1}\hat Z_K^{-l}
\bar\Gamma_{2+2n,l}\;.
}
We point out that these redefinitions correspond to the imposition 
of the zero-momentum renormalization prescriptions for any $\Lambda$,
\formula{norm}
{\prendi{\partial_{p^2}\hat\Gamma_2}{p=0}\equiv1,\quad
\prendi{\hat\Gamma_2}{p=0}\equiv\Lambda^2,\quad\prendi{\hat\Gamma_4}{p_i=0}
\equiv\hat\lambda(\Lambda),\quad\prendi{\hat\Gamma_{2,1}}{p_i=0}\equiv1.
}
The left hand side of the evolution equation 
for the rescaled functional
$\hat\Gamma(\hat\phi,\hat K,\hat\lambda;\Lambda)$ reads
\formula{eq.riscalata}
{\LdL\Gamma=
\left(\LDL+\frac12\frac{\LdL\hat Z_\phi}{\hat Z_\phi}
\hat\phi\cdot\dd{}{\hat\phi}
+\frac{\LdL Z_K}{Z_K}K\cdot\dd{}K+\LdL\hat\lambda\dede{}{\hat\lambda}\right)
\hat\Gamma,
}
where $\Lambda\delta_\Lambda$ denotes 
the partial derivative with respect to the explicit
$\Lambda-$depen\-dence of $\hat\Gamma(\hat\phi,\hat K,\hat\lambda;\Lambda)$.
It is also convenient to define
\formula{beta.gamma}
{\hat\gamma_\phi=-\frac12\frac{\LdL\hat Z_\phi}{\hat Z_\phi},\quad
\hat\gamma_K=-\frac{\LdL \hat Z_K}{Z_K},\quad\hat\beta\equiv\LdL{\hat\lambda}.
}
With these notations the left hand side of the evolution equation
on proper vertices reads
\formula{LHS.pv}
{\LdL\Gamma_{2n,l}(p_i,q_j;\Lambda)=
\left(\LDL-2n\hat\gamma_\phi-l\hat\gamma_K+\hat\beta
\dede{}{\hat\lambda}\right)\hat\Gamma_{2n,l}(p_i,q_j;\Lambda).
}
In order to recover the Callan-Symanzik equation we observe that
in the case of the mass cutoff $(\hat Q_\Lambda=\Lambda^2)$ the relation
$\LdL Q_\Lambda=2(1-\hat\gamma_\phi)Q_\Lambda$
holds; therefore using $\LdL\Gamma=-\LdL \ln Z$ 
and the path integral representation of the
partition function $Z(J,K;\Lambda)$ (suitably regularized in the
ultraviolet) the right hand side of the
evolution equation can be written as
\formula{RHS}
{-\frac{\LdL Z}Z=\int_x \frac12\LdL Q_\Lambda<\phi^2(x)>_{J,K}=
(1-\hat\gamma_\phi)
\Lambda^2<\hat\phi^2(x)>_{J,K}.
}
Replacing now the expectation value $\Lambda^2<\hat\phi^2(x)>_{J,K}$ with
$-2\Lambda^2\int_x\dd {\hat W}{\hat K(x)}=2\Lambda^2\int_x\dd{\hat\Gamma}
{\hat K(x)}$ we immediately see that
the Wilson evolution equation on proper vertices assumes
the textbook Callan-Symanzik form \cite{ZJ.book}
\formula{CS.pv}
{\left(\LDL-2n\hat\gamma_\phi-l\hat\gamma_K+\hat\beta
\dede{}{\hat\lambda}\right)\hat\Gamma_{2n,l}=
2\Lambda^2(1-\hat\gamma_\phi)\hat\Gamma_{2n,l+1}(p_i,q_j,0;\Lambda)\;.
}
An alternative more explicit form for \rif{CS.pv} is
\formula{WCS}
{\left(\LDL-\hat\gamma_\phi{\hat\phi}\cdot\dd{}{\hat\phi}-
\hat\gamma_K\hat K\cdot\dd{}{\hat K}+\hat\beta
\dede{}{\hat\lambda}\right)\hat\Gamma=
\left(1-\hat\gamma_\phi\right)\hat I(\hat\phi,\hat K;\hat\lambda;\Lambda)}
with 
\formula{hat.I}
{\hat I(\hat\phi,\hat K;\hat\lambda;\Lambda)=
\int_x \Lambda^2\hat\phi^2(x)-\int_{xyz}\Lambda^2
\hat\Gamma_2^{-1}(x-y)\hat{\bar\Gamma}_{\phi\phi}(y,z)\hat\Gamma_2^{-1}(z-x).}
We can consider separately the relevant and the irrelevant part of 
equation \rif{WCS}. 
\begin{itemize}
\item
The relevant part is described by the
functions $\hat\gamma_\phi,\hat\gamma_K$ and $\hat\beta$. In fact, since 
the rescaling factors were chosen in such a way to canonically normalize
the kinetic term and the $\hat\phi^2/2$ insertion
we can compute $\hat\gamma_\phi$ and $\hat\gamma_K$ from 
\rif{WCS} and \rif{norm}, by obtaining 
\formula{Z2}
{\hat\gamma_\phi=\frac{-\frac12\partial_{p^2}\hat I_2|_0}
{1-\frac12\partial_{p^2}\hat I_2|_0}
}
and
\formula{gamma.K}
{\hat\gamma_K=-(1-\gamma_\phi)\prendi{\hat I_{2,1}}{p_i=0}-2\gamma_\phi.
}
Moreover $\hat\beta$ can be computed
from the four point vertex,
\formulaX
{\hat\beta=4\hat\gamma\hat\lambda+(1-\hat\gamma)\hat I_4|_{p_i=0}.
} 
It is a simple 
exercise to compute $\hat\gamma_\phi,\ \hat\gamma_K$ and $\hat\beta$ at the
lowest order in perturbation theory, obtaining the usual results
\formulaX{
\hat\gamma_\phi^{(1)}=0,\quad\hat\gamma_K^{(1)}=
-\frac1{16\pi^2}\hat\lambda,\quad\hat\beta^{(1)}=\frac3{16\pi^2}\hat\lambda^2.}
\item The irrelevant part  of the evolution equation \rif{WCS}
in the critical regime 
$\Lambda^2<<p_i^2$ can be simply neglected.
In fact, by dimensional analysis, we see that the irrelevant (i.e.
zero-momentum subtracted) vertices 
are suppressed as inverse powers of the momenta. Therefore in this regime
the usual asymptotic Callan-Symanzik equation \cite{CS} holds
\formula{CS}
{\left(\LDL-2n\hat\gamma_\phi-l\hat\gamma_K+\hat\beta
\dede{}{\hat\lambda}\right)\hat\Gamma_{2n,l,irr}(p_i,\hat\lambda;\Lambda)
\simeq0,\quad\frac{\Lambda^2}{p_i^2}\to0.
}
\end{itemize}

The deduction of the scaling equation \rif{CS} from the Exact Renormalization
Equation,
even if particularly transparent in our approach, is actually quite general
and independent of the technical implementation of the evolution equation.
In fact, in the asymptotic regime the right hand side 
of equation \rif{CS} vanishes by power counting independently of 
the specific form of the evolution equation. Moreover the $\hat\gamma_\phi,
\hat\gamma_K$ and $\hat\beta$ functions are universal at first order 
in perturbation theory (actually the beta function is universal up to the
second order in perturbation theory \cite{ZJ.book}).

We point out that an extension to gauge theories 
is straightforward only within our gauge-consistent formalism. 
Otherwise we are forced to follows the running of spurious
couplings, related to the non-gauge-invariant operators,
in terms of the physical one's
with the cumbersome mechanism of broken Ward identities or
fine-tuning conditions. 
As a matter of fact, at one-loop the spurious
couplings are finite in the large$-\Lambda$ limit (except the mass coupling
which is quadratically divergent) therefore in practice in this approximation
a similar approach is
suitable also with generic cutoffs \cite{BMS,BS}. Nevertheless, 
at higher loops, also the
spurious couplings develop logarithmic divergences, 
even if they are sub-leading as a consequence
of the (broken) Ward identities, and should be considered. All these 
complications are avoided in our gauge-invariant formulation and one
obtains the expected asymptotic
Callan-Symanzik equation for QED in the critical regime
\formula{QED.crit}
{m^2<<\Lambda^2<<p_i^2<<\Lambda_{UV}^2
}
i.e. when $\Lambda$ is large with respect to $m$ but small with respect
to the momenta and the ultraviolet cutoff. The explicit formula is the
usual one \cite{CS}
\formula{QED.RG}
{(\Lambda\delta_\Lambda-\hat\gamma_A N_A -\hat\gamma_\psi(N_\psi+N_{\bar\psi})+
\hat\beta_e\partial_{\hat e}+\hat\beta_\xi\partial_{\hat\xi})\hat\Gamma_{irr}
(p_i;\Lambda)\buildrel{\Lambda\to0}\over=0
}
where
\formulaX
{N_A=\int_xA_\mu\dd{}{A_\mu},\quad N_\psi=\int_x\psi\dd{}{\psi},
\quad N_{\bar\psi}=\int_x\bar\psi\dd{}{\bar\psi}
}
and, as a consequence of Ward identities,
\formulaX
{\hat e=e/\hat Z_A^{1/2},\quad\hat\xi=\xi \hat Z_A,\quad
\hat\beta_e=\hat e\hat\gamma_A,\quad\hat\beta_\xi=-2\hat\xi\hat\gamma_A.
}
We stress that 
in our approach all the usual consequence of gauge-invariance apply and 
in particular the unitarity property holds. 
Moreover one has the usual 
control on the gauge-parameter dependence \cite{ZJ.book} and can prove that
the flowing beta function is gauge-independent to all orders. On the contrary
all these important properties do not have any simple analytic control with 
generic Wilsonian procedures. Equation \rif{QED.RG} can be also
seen as a starting point for an improved perturbation theory in the sense 
of \cite{BMS,BS}.

Finally, we remind the reader 
that there is a one-to-one correspondence between 
the Wilson or Callan-Symanzik renormalization group functions 
$\hat\beta$ and $\hat\gamma$ 
and the corresponding 
$\beta$ and $\gamma$ of the Gell-Mann and Low renormalization group, obtained
by imposing the independence of the bare (ultraviolet) objects from the 
renormalization point $\mu$. The interested reader is referred to \cite{BDM} 
for the general formulae 
and some explicit computation
in YM theories (at one-loop) and in $\phi^4_4$ (at two-loops).
 
\section{Conclusions and outlook}

In this paper we succeeded to give an explicitly gauge consistent Wilson
Renormalization Group formulation of Quantum Electrodynamics. 
The solution of the problem is
based on a specific choice of the infrared cutoff,  
corresponding to a mass term for both the photon and the electron, supplemented
with a gauge-invariant ultraviolet regularization. 
In this context the Callan-Symanzik equation is equivalent to the 
Wilson's equation and can be used in the study
both of perturbative and non-perturbative applications. 

On the perturbative side, a possible application of the scheme we propose
is in the problem of the renormalization of composite operators
in gauge theories. Here the mass cutoff is very convenient since it avoids
the problem of the usual Wilsonian approaches where gauge-invariance 
is broken and gauge-invariant operators unavoidably 
mix with non-gauge invariant operators. Actually, 
in the issue of the perturbative computation of anomalous
dimensions the formalism we present has the same level of efficency of 
dimensional regularization.

The most important application of the Wilsonian approach
is in the study of non-perturbative aspects.
There are various well studied numerical methods of solution of the ERGE  
in the literature, 
based on some truncations of the effective action \cite{Wetterich1} 
or the derivative expansion \cite{MorrisTrunc}. The important point is that
since our version of the ERGE is gauge-consistent, 
no gauge-variant terms are generated by the evolution, as instead 
happens in generic Wilsonian approaches \cite{giap.QED}.
In general, non-perturbative methods can be applied to our formulation,
provided that we renormalize correctly the theory in the ultraviolet.
We also mention that there is a paper in the literature 
in which the mass cutoff is introduced and used in order
to study non-perturbative aspects
of the $\phi^4$ theory at next to leading order in the gradient expansion
\cite{Polonyi}.

There are a number of possible extensions of this work to other theories.
\begin{itemize}

\item A trivial extension is the application to scalar QED. The procedure
works exactly as in the spinor case, provided that we use the following cutoff
function for the scalars:
\formula{scal.cutoff}
{\tilde\kIR(q)=\frac{q^2-m^2}{q^2-m^2-\Lambda^2}.
}
In this way the quadratic term $\phi^*\tilde Q_\Lambda\phi=
\Lambda^2\phi^*\phi$ is explicitly gauge invariant and the 
Ward identities breaking is exactly as in \rif{Ward.gen}.
The  three-dimensional case has been studied non-perturbatively
in \cite{Freire} but without control of gauge-invariance and it would be
a very practical model to test the method since no
ultraviolet regularization is needed.

\item Another straightforward extension is the application 
to supersymmetric Abelian gauge 
theories, because the Wilsonian formulation is consistent with
supersymmetry \cite{BV.super,Geyer}. In this case it is sufficient to add to
the usual classical gauge-invariant 
action of super QED  $S_{CL}(V,\phi_\pm,\phi^\dagger_\pm)$
the quadratic term
$$
\int_{x,\theta,\bar\theta} \left[-\frac1{32}\Lambda^2 V^2-\frac1{128\xi}
D^2 V\bar D^2 V\right]+\frac1{32}\int_{x,\theta}i\Lambda\phi_+\phi_- +
\mbox{h.c.}
$$
In this way $\Gamma^{(0)}(V,\phi_\pm,\phi_\pm^\dagger;\Lambda)$ 
is well behaviored under the infinitesimal gauge 
transformation
$$
\W_f V=f+f^*,\quad \W_f \phi_\pm=\pm ie f\phi_\pm,\quad \W_f\phi^\dagger_\pm=
\mp ie f\phi^\dagger_\pm
$$
in the sense that the breaking term is linear in the
field $V$, i.e. we can maintain to all orders the Ward identity
$$
\W_f\Gamma=\int_{x,\theta,\bar\theta}
(f+f^*)\left(-\frac1{16}\Lambda^2+\frac{\bar D^2 D^2+D^2\bar D^2}{128\xi}
\right)
$$
which is the supersymmetric generalization of \rif{Ward.gen}.
We remark that even the analysis of supersymmetric 
theories is particularly simple since in this case 
mass divergences automatically cancel and therefore the
step of an intermediate regularization
can be avoided. This is ultimately related to the supersymmetric solution  of
the naturalness problem \cite{Polchinski} i.e. the absence of quadratic
divergences.

\item From these examples it is clear that our procedure formally works
for any theory characterized by linearly broken Ward identities. 
In particular the formalism applies to non-Abelian gauge theories in algebraic
non-covariant
gauges. However, in this case, one expects some difficulty related to the
presence of a gluon ``mass'' $\Lambda^2\neq0$. 
A detailed study of the question is given in \cite{paper.II}.

\end{itemize} 

\vskip .5cm
{\large\bf Acknowledgements}\\

I thank M. Bonini for reading a first version of the manuscript, 
foundation ``Aldo Gini'' and INFN, Gruppo Collegato di Parma 
for financial support in the early stages of this work.

\appendix

\section{The exact evolution equation}

For sake of completeness, and in order to fix the notation, 
here we briefly review the deduction of the ERGE in the formalism of
the cutoff effective action $\Gamma(\Lambda)$. The standard
deduction can be easily generalized in order to manage the problem of
the renormalization of composite operators, simply by introducing sources $K$
associated with the operators of interest. The simplest case, as seen
in section 6, is the operator $\frac12\phi^2(x)$ in the $\phi^4$ theory.
In the following, for sake of convenience,
we directly work in the Minkowski space
even if an analytic continuation in the Euclidean space should be understood 
in order to give a rigorous meaning to the momentum integrals, and also 
to have a more clear Wilsonian interpretation.

The starting point is the cutoff generating functional
\formula{Z.Lambda}
{Z(J,K;\Lambda)=\int[d\Phi]e^{iS_B(\Phi,K)+\frac i2\tilde\Phi Q_\Lambda\Phi+
i J\cdot\Phi}
}
in which, since the quadratic term cuts the infrared modes, in practice
we are integrating only the degrees of freedom {\it over} $\Lambda$,
i.e. the ultraviolet modes. 

In equation \rif{Z.Lambda} $S_B(\Phi,K)$ denotes
the bare action and some kind of ultraviolet regularization is understood
even if it is not explicitly displayed. 
The evolution equation is derived simply by studying the behavior of
the generating functional under variation of the infrared cutoff $\Lambda$.
One readily obtains
\formula{evoluzione Ze}
{\LdL Z(J,K;\Lambda)=\frac i2(-)^A \dot 
Q_{\Lambda,AB}\dd{}{iJ_B}\ddes{}{iJ_A} Z(J,K;\Lambda),
}
where the de Witt notation is used, i.e. the indices $A,B$ represents both
continuous and discrete indices and sums and 
and integrals are understood. The symbol $(-)^A$ gives a plus
sign for bosonic fields and a minus sign for fermionic fields.
The functional derivative operator $\dd{}{iJ_B}$ acts from the left
whereas $\ddes{}{iJ_B}$ acts from the right. We also
use the abbreviations
\formulaX
{\tilde\Phi Q_\Lambda\Phi=(-)^A\Phi^A Q_{\Lambda,AB}\Phi^B,\quad\tilde \Phi^A=
(-)^A\Phi^A.}
From equation \rif{evoluzione Ze} one can obtain
the evolution of the generating functional of cutoff connected
Green functions $W(J;\Lambda)=-i\log Z(J;\Lambda)$,
\formula{evoluzione W.gen}
{\LdL W=-\frac i2(-)^A \dot Q_{\Lambda,AB}\left(\dd{}
{J_B}\ddes{}{J_A}W+i\dd{W}{J_B}\ddes {W}{J_A}\right).
}
We recall that $W(J,0;\Lambda)$ is directly 
related to the Wilsonian effective action \cite{BDM,Morris 1}.
Here we are interested in
the evolution equation for the Legendre transformed effective action
(simply called cutoff effective action or $\Lambda-$RG action)
\formula{Gamma.def}
{\Gamma(\Phi,K;\Lambda)=-J_A\Phi^A+W(J,K;\Lambda),\quad\Phi^A=\dd W{J_A}.
}
With some simple manipulation one obtains
\formulona{ev.Gamma.gen}
{\LdL\left(\Gamma-\frac12\tilde\Phi Q_\Lambda\Phi\right)
=I(\Phi,K;\Lambda)=\frac i2\STr\dot Q_\Lambda\Gamma_{\tilde\Phi\Phi}^{-1}
}
where the supertrace notation is used and we have defined
\formulaX
{(\Gamma_{\tilde\Phi\Phi}^{-1})^{BA}\equiv-\left.\dd{}
{J_B}\ddes{}{J_A}W\right|_{J=J(\Phi)}.
}
Although this form of the evolution
equation is well suited for non-perturbative studies, in particular
to perform truncation of the evolution equation
such for instance the Local Potential Approximation 
\cite{Wetterich1,MorrisTrunc}, nevertheless in order to extract the
perturbative expansion and to give renormalizability proofs
another form of \rif{ev.Gamma.gen} is more convenient. To this aim we
introduce an auxiliary functional $\bar\Gamma_{\Phi\Phi}$, implicitly
defined by the relation
\formula{invers.Gamma}
{\Gamma^{-1}_{\tilde\Phi\Phi}(\Phi,K;\Lambda)=\Gamma_2^{-1}-
\Gamma_2^{-1}\bar\Gamma_{\tilde\Phi\Phi}(\Phi,K;\Lambda)\Gamma_2^{-1},
}
where $\Gamma_2\equiv\Gamma_{\tilde\Phi\Phi}|_{\Phi=0,K=0}$ is the two-point
function.
Now the right hand side of equation \rif{ev.Gamma.gen} can be rewritten in 
the condensed form
\formula{def.I}
{I(\Phi,K;\Lambda)\equiv-\frac i2\STr\dot Q_\Lambda
\Gamma_2^{-1}(\bar\Gamma_{\tilde\Phi\Phi}\Gamma_2^{-1}-1).
}
It is convenient to introduce the following condensed notation for the proper
vertices without insertions of operators ($K=0$)
\formulona{proper.vertices}
{\Gamma(\Phi;\Lambda)=\frac1{n!}\sum_{n=2}^\infty
\Phi^{A_n}\dots \Phi^{A_1}\Gamma_{A_1\dots A_n}
\\
I(\Phi;\Lambda)=\frac1{n!}\sum_{n=2}^\infty\Phi^{A_n}\dots 
\Phi^{A_1}I_{A_1\dots A_n}
\\
\bar\Gamma_{\tilde\Phi\Phi,AB}(\Phi;\Lambda)=\frac1{n!}\sum_{n=1}^\infty
\Phi^{A_n}\dots \Phi^{A_1}\bar\Gamma_{AA_1\dots A_nB}
}
(for instance   $\Gamma_{A_1A_2}$ in a more explicit notation
corresponds both to $\Gamma_{\mu\nu}(p;\Lambda)\cdot\\ 
(2\pi)^4\delta^4(p+q)$ and to 
$\Gamma_{\alpha\beta}(p;\Lambda)(2\pi)^4\delta^4(p+q)$). 
In this way the auxiliary functional $\bar\Gamma_{\tilde\Phi\Phi}$ 
introduced in \rif{invers.Gamma}
can be explicited by using the recursive formula
\formula{bar.Gamma.n.gen}
{\bar\Gamma_{AA_1\ldots A_nB}=\Gamma_{AA_1\ldots A_nB}-\sum_{k=1}^{n-1}
\Gamma_{AA_1\dots A_k C}(\Gamma_2^{-1})^{CD}\bar\Gamma_{DA_{k+1}\ldots A_nB}.
}
The graphical representation of equation \rif{bar.Gamma.n.gen} is reported in
figure 1. See also \cite{BDM,BDM.QED} for explicit examples.
The vertices with insertions of operators are obtained straightforwardly
by deriving successively the evolution equation with respect to the sources $K$
and by taking $K=0$.

\section{Explicit form of relevant and irrelevant functionals for QED}

In the QED case the relevant effective action has the following 
Ward-iden\-tities-consistent form,
\formulaX
{\Gamma_{rel}(\Phi;\Lambda)=
\int_x\frac12\Lambda^2A\cdot A-\frac1{2\xi}
(\partial\cdot A)^2+\Gamma'_{rel}(\Phi;\Lambda),}
where
\formulonaX
{\Gamma'_{rel}(\Phi;\Lambda)&=
\int_k-\frac12c_A(\Lambda)A_\mu(-k) k^2 t^{\mu\nu}(k) A_\nu(k)
+\int_x e\ c_\psi(\Lambda)\bar\psi\slash A\psi\cr
&\quad\int_p\bar\psi(p)\left[\slash p c_\psi(\Lambda)-c_1(\Lambda)-
ic_2(\Lambda)\gamma_5\right]\psi(p),
}
For sake of brevity, we omitted the analogous contributions
for Pauli-Villars fields and we defined
\formula{t.l}
{t_{\mu\nu}(k)=g_{\mu\nu}-\ell_{\mu\nu}(k),\quad \ell_{\mu\nu}(k)
=k_\mu k_\nu/k^2.  
}
The relevant couplings are 
\formule{c_QED}
{c_A(\Lambda)&=&
-\prendi{\partial_{k^2}\frac13 t^{\mu\nu}\Gamma_{\mu\nu}}{k=0},\\
c_\psi(\Lambda)&=&\prendi{\partial_{p^\mu}\frac1{16} \gamma^{\mu,\beta\alpha}
\Gamma_{\alpha\beta}}{p=0},\\
c_1(\Lambda)&=&-\frac14\prendi{\delta^{\beta\alpha}\Gamma_{\alpha\beta}}{p=0},\\
c_2(\Lambda)&=&\prendi{\frac i4
\gamma_5^{\beta\alpha}\Gamma_{\alpha\beta}}{p=0},
}
where $\Gamma_{\mu\nu}$ and $\Gamma_{\alpha\beta}$ are the photon and electron 
two-point functions.
The renormalization prescriptions are
\formula{ren.pr.QED}
{c_A(\Lambda_R)=1,\quad c_\psi(\Lambda_R)=1,\quad c_1(0)=m,\quad c_2(0)=0.
}
The irrelevant part of the photon two-point function is given by the formula
\formula{ph.pr}{\Gamma_{\mu\nu,irr}(k;\Lambda)=\Gamma_{T,irr}(k^2;\Lambda)\ 
t_{\mu\nu}(k)+\Gamma_{L,irr}(k^2;\Lambda)\ \ell_{\mu\nu}(k),
}
where
\formuleX
{\Gamma_{L,irr}(k;\Lambda)&=&\Gamma_L-
k^2\partial_{\bar k^2}\Gamma_L|_{\bar k=0}-\Gamma_L|_{\bar k=0},\\
\Gamma_{T,irr}(k;\Lambda)&=&\Gamma_T-
k^2\partial_{\bar k^2}\Gamma_T|_{\bar k=0}-\Gamma_T|_{\bar k=0}.
}
The irrelevant part of the electron two-point function is
$$
\Gamma_{\alpha\beta,irr}(p)=\Gamma_{\alpha\beta}(p)-
\prendi{\Gamma_{\alpha\beta}(\bar p)}{\bar p=0}-
p_\mu\prendi{\frac{\partial}{\partial \bar p_\mu}
\Gamma_{\alpha\beta}(\bar p)}{\bar p=0}.
$$
The irrelevant part of the photon-electron-positron vertex 
is
$$
\Gamma_{\mu\alpha\beta,irr}(p,p')=\Gamma_{\mu\alpha\beta}(p,p')-
\Gamma_{\mu\alpha\beta}(0,0).
$$
The same decomposition into relevant and irrelevant parts holds 
for the vertices of the functional $ I(\Phi;\Lambda)$ 
and also in the rescaled case, 
i.e. for the functionals $\hat\Gamma(\hat\Phi;\hat\lambda;\Lambda)$ and 
$\hat I(\hat\Phi;\hat\lambda;\Lambda)$.

As we remarked in section 6, 
to properly renormalize the theory one should also consider the 
renormalization of the composite gauge-invariant operators 
$O_1(x)=\bar\psi\psi$
and $O_2(x)=i\bar\psi\gamma_5\psi$, which can be performed straighforwardly
by adding the corresponding external sources $K_1(x)$ and $K_2(x)$.
This gives two new dimensionless relevant couplings $\sigma_1(\Lambda)$ and
$\sigma_2(\Lambda)$ and two new renormalization prescriptions and subtractions.
We still stress that differently from other Wilsonian procedures 
spurious couplings corresponding
to non-gauge-invariant operators are never generated.

\section{The renormalizability proof} 

In this section we give a simple renormalizability proof  which generalizes
the analysis of \cite{BDM} to a quite large
class of cutoff functions. 
For notational commodity we first present the method for the massless
euclidean $\phi^4_4$ theory. We denote by $M_0$ the mass scale where
the ultraviolet regularization becomes effective and we 
give the proof for cutoff functions such as the integral
\formula{utile}
{\lim_{M_0\to\infty}
\int_q|\dot \Delta_{\Lambda\infty}(q;M_0)\Delta^{-1}_{\Lambda\infty}(q;M_0)|=
c\Lambda^4
}
is finite. 
For instance the exponential cutoff $K_{\Lambda\infty}(q)=
1-\exp(-q^2/\Lambda^2)$ satisfies \rif{utile} with coefficient
$c=\zeta(3)/(4\pi^2)$. Notice that the mass cutoff does not belong 
to this class and the renormalizability proof requires 
a different analysis \cite{Callan}. For this reason we prefer to
give the proof in the additional hypothesis \rif{utile},
which is not needed from a rigorous point of view, but 
it is technically very convenient.

The renormalizability proof
is essentially based on the perturbative evolution equation 
\rif{eq.int.2n} which gives the proper vertices at loop $\ell+1$ 
in terms of integrals containing the proper vertices at lower loops. 
We simply prove that these integrals are well defined when the ultraviolet
regularization is removed, i.e. $M_0\to\infty$. Notice that
the infrared behavior is safe by construction, 
because the infrared cutoff $\Lambda$ at this level is assumed non-zero.
It is convenient to define the norms at loops $\ell'=0,1,2,\dots$,
\formula{norma1}
{||\Gamma^{(\ell')}_{2n}||_\Lambda\equiv
\lim_{M_0\to\infty}
\max_{p_i^2<\Lambda^2}|\Gamma^{(\ell')}_{2n}(p_1\ldots p_{2n};\Lambda,M_0)|.
}
The tree level vertices have finite norm since
$||\Gamma^{(0)}_{2n}||_\Lambda\sim\Lambda^{4-2n}$ is finite for
$\Lambda\neq0$.
It is also convenient to introduce the functions
$$X_{2n+2}^{(\ell')}=\frac12
[\Gamma_2^{-1}(q;\Lambda,M_0)
\bar\Gamma_{2n+2}(q,p_i,-q,\Lambda,M_0)
\Gamma_2^{-1}(q;\Lambda,M_0)\Delta_{\Lambda\infty}^{-1}(q;M_0)]^{(\ell')}$$ 
and their norms
\formula{norma2}
{||X^{(\ell')}_{2n+2}||_\Lambda\equiv
\lim_{M_0\to\infty}\max_q\max_{p_i^2<\Lambda^2}
|X^{(\ell')}_{2n+2}(q,p_i,-q;\Lambda,M_0)|.
}
At tree level $||X^{(0)}_{2n+2}||_\Lambda\sim\Lambda^{-2n}$ 
is finite for $\Lambda\neq0$. In general, if the norms
$||\Gamma^{(\ell')}_{2n}||_\Lambda$ are finite for all loops $\ell'\leq\ell$,
then the norms  $||X^{(\ell)}_{2n+2}||_{\Lambda}$ are finite,
since they are obtained from functions $\Gamma^{(\ell')}_{2m}$ 
with $2m\leq 2n+2$ and $\ell'\leq\ell$,
by using the recursive relation \rif{bar.Gamma.n}
between functions $\bar\Gamma_{2n+2}$ and vertices $\Gamma_{2m}$.
With these notations the evolution equation reads
\formula{PRGE}
{\dot\Gamma_{2n}^{(\ell+1)}=\int_q
\dot \Delta_{\Lambda\infty}(q;M_0)\Delta^{-1}_{\Lambda\infty}(q;M_0)
X^{(\ell)}_{2n+2}(q,p_i,-q;\Lambda,M_0).
}
We split the renormalizability proof in four steps.

\begin{enumerate}
\item Inductive hypothesis at loop $\ell$.
Due to Lorentz-invariance the proper vertices
$\Gamma_{2n}(p_i;\Lambda)$ only
depend of the invariant combinations $s_k=(p_i^2,p_i\cdot p_j)$ (there are
$n(2n-1)$ independent invariants). We take as inductive hypothesis\footnote
{Actually we expect some logarithmic behavior, and a better \ansatz should be
$$
||\Gamma_{2n}^{(\ell')}||_\Lambda=\Lambda^{4-2n}
P_{2n}^{(\ell')}\left(\log\frac{\Lambda^2}{\Lambda_R^2}\right),
$$ where $P_{2n}^{(\ell')}$ is a polynomial of degree increasing with the loop
number $\ell'$. However this does not change
our conclusions about the convergence of integrals. One can easily
prove that this
\ansatz is consistent with the evolution equation, i.e. assuming the \ansatz 
at loop $\ell$, it holds at loop $\ell+1$.
} 
\formula{ip.ind}
{||\partial_{s_{k_1}}\dots\partial_{s_{k_m}}
\Gamma^{(\ell')}_{2n}||_\Lambda\sim \Lambda^{4-2n-2m}<\infty,\quad
\ell'=0,\dots,\ell.}
From \rif{ip.ind} we have
\formulaX
{||\partial_{s_{k_1}}\dots\partial_{s_{k_m}}
X^{(\ell')}_{2n+2}||_\Lambda\sim \Lambda^{-2n-2m}<\infty\,\quad
\ell'=0,\dots,\ell .}

\item Lemma 1.
As a direct consequence of property \rif{utile} and
definitions \rif{norma1},\rif{norma2}, the inequality
\formulaX
{||\int_q\dot \Delta_{\Lambda\infty}(q)\Delta^{-1}_{\Lambda\infty}(q)
X^{(\ell')}_{2n+2}
(q,p_1\ldots p_{2n},-q)||_\Lambda
\leq c\Lambda^4||X^{(\ell')}_{2n+2}||_\Lambda
}
holds.
\item Lemma 2.
There are important bounds for the irrelevant vertices
$\Gamma^{(\ell')}_{2,irr}$ and $\Gamma^{(\ell')}_{4,irr}$. 
As a consequence of the identities
\formulaX
{f(z)-f(0)-zf'(0)=z^2\int_0^1 dx (1-x)f''(z x)}
and
\formulaX
{f(z)-f(0)=z\int_0^1 dx f'(z x),}
which hold for any analytic function, the inequalities
\formula{g2.der}
{||\Gamma^{(\ell')}_{2,irr}||_\Lambda=||\Gamma^{(\ell')}_2(s)-
\Gamma^{(\ell')}_2(0)-s\partial_s\Gamma^{(\ell')}_2(0)||_\Lambda\leq
\frac{\Lambda^4}2||\partial_s^2\Gamma^{(\ell')}_2||_\Lambda
}
and
\formula{g4.der}
{||\Gamma^{(\ell')}_{4,irr}||_\Lambda
=||\Gamma^{(\ell')}_4(p_i)-\Gamma^{(\ell')}_4(0)||_\Lambda\leq
\Lambda^2||\partial_{s_k}
\Gamma^{(\ell')}_4||_\Lambda
}
hold.
\item Inductive hypothesis at loop $\ell+1$.
We have to prove
\formulaX
{||\Gamma_{2n}^{(\ell+1)}||_\Lambda\leq||\Gamma_{2n,rel}^{(\ell+1)}||_\Lambda
+||\Gamma_{2n,irr}^{(\ell+1)}||_\Lambda<\infty.}
The finiteness of $||\Gamma_{2n,rel}^{(\ell+1)}||_\Lambda$, i.e. of
relevant coefficients, comes from the inductive hypothesis
and lemma 1:
\formuleX
{|c_m^{(\ell+1)}(\Lambda)|&\leq&
\int_0^\Lambda\frac{d\Lambda_\ell}{\Lambda_\ell}
\left|\int_q\dot \Delta_{\Lambda_\ell\infty} \Delta^{-1}_{\Lambda_\ell\infty}
X_4^{(\ell)}|_0\right|\sim \Lambda^2\\
|c_\phi^{(\ell+1)}(\Lambda)|
&\leq&\int_{\Lambda_R}^\Lambda\frac{d\Lambda_\ell}{\Lambda_\ell}
\left|\int_q\dot \Delta_{\Lambda_\ell\infty}\Delta^{-1}_{\Lambda_\ell\infty}
\partial_s X_4^{(\ell)}
|_0\right|\sim \Lambda^0\\
|c_\lambda^{(\ell+1)}(\Lambda)|&\leq&
\int_{\Lambda_R}^\Lambda\frac{d\Lambda_\ell}{\Lambda_\ell}
\left|\int_q\dot \Delta_{\Lambda_\ell\infty} \Delta^{-1}_{\Lambda_\ell\infty}
X_6^{(\ell)}|_0\right|\sim \Lambda^0.}
For the irrelevant vertices by using lemma 1 and lemma 2 we obtain
\formulonaX
{||\Gamma_{2,irr}^{(\ell+1)}||_\Lambda&\leq\frac{\Lambda^4}2\int_\Lambda^\infty
\frac{d\Lambda_\ell}{\Lambda_\ell} ||\int_q\dot \Delta_{\Lambda_\ell\infty} 
\Delta^{-1}_{\Lambda_\ell\infty}\partial_s^2 X_4||_{\Lambda_\ell}\\
&\leq\frac{c\Lambda^4}2\int_{\Lambda}^\infty
\frac {d\Lambda_\ell}{\Lambda_\ell} \Lambda_\ell^4\ \Lambda_\ell^{-6}\sim 
\Lambda^2
}
and analogously
\formulaX
{||\Gamma_{4,irr}^{(\ell+1)}||_\Lambda\leq\Lambda^2\int_{\Lambda}^\infty
\frac {d\Lambda_\ell}{\Lambda_\ell} ||\int_q\dot \Delta_{\Lambda_\ell\infty} 
\Delta^{-1}_{\Lambda_\ell\infty}\partial_{s_k} X_6||_{\Lambda_\ell}\sim 
\Lambda^0}
\formulaX
{||\Gamma_{2n}^{(\ell+1)}||_\Lambda\leq\int_{\Lambda}^\infty
\frac {d\Lambda_\ell}{\Lambda_\ell}
||\int_q\dot \Delta_{\Lambda_\ell\infty}\Delta^{-1}_{\Lambda_\ell\infty} 
X_{2n+2}||_{\Lambda_\ell}\sim \Lambda^{4-2n}.
}
The convergence of $\Lambda_\ell-$integrals is guaranteed for the power 
counting and the subtractions \rif{g2.der},\rif{g4.der}; therefore
the proper vertex at loop $\ell+1$ are well defined. 
{\it A fortiori} that holds for the derivatives
$||\partial_{s_{k_1}}\dots\partial_{s_{k_m}}\Gamma^{(\ell+1)}_{2n,irr}
||_\Lambda<\infty$. Therefore the inductive hypothesis \rif{ip.ind}
holds at loop $\ell+1$ also for irrelevant vertices. By induction,
the renormalizability proof holds 
to any {\it finite} order $\ell$.

\end{enumerate}

The same approach can be applied to the QED case:
one easily prove that all the $\Lambda_\ell-$integrals 
are well defined by
using the subtractions and the behavior expected by dimensional analysis,
\formula{behavior.QED}
{
\lim_{M_0\to\infty}\max_{p_i^2\leq \Lambda^2}
|\Gamma_{n_{\bar\psi\psi},n_A}^{(\ell')}(p_i;\Lambda)|\sim
\Lambda^{4-3n_{\bar\psi\psi}-n_A},
}
where $n_A$ is the number of external photon lines and $n_{\bar\psi\psi}$
the number of external fermion-antifermion lines.

Actually, 
one can easily convince himself that this kind of proof holds for any
theory respecting the power counting criterium.

\end{document}